\documentclass[aps,prd,notitlepage,twocolumn,
nofootinbib]{revtex4-1}

\usepackage{amsmath, amsthm, amssymb, mathtools, bm, mathrsfs, slashed, relsize}

\usepackage{epsfig,color,graphicx}
\usepackage{xcolor}
\usepackage{subfigure}
\colorlet{darkred}{red!80!black}
\colorlet{darkgreen}{green!50!black}
\colorlet{darkblue}{blue!50!black}
\colorlet{darkviolet}{violet!50!black}

\usepackage{soul}

\usepackage[pdftex,hypertexnames=false,linktocpage=true]{hyperref}
\hypersetup{colorlinks=true,linkcolor=blue,anchorcolor=blue,citecolor=blue,filecolor=blue,urlcolor=black,bookmarksnumbered=true,
pdfview=FitB
}

\allowdisplaybreaks

\mathchardef\-="2D
\def\Tr{{\textbf{Tr}}}

\newcommand{\cd}{\makebox[0.08cm]{$\cdot$}}

\newcommand{\eg}{\textit{e.g.}}  

\newcommand{\ie}{\textit{i.e.}}

\DeclareMathAlphabet{\mathpzc}{OT1}{pzc}{m}{it}

\begin{document}

\title{Helium-3 relativistic wave function in light-front dynamics}


\author{Zhimin Zhu}
\email{zhuzhimin@impcas.ac.cn}
\affiliation{Institute of Modern Physics, Chinese Academy of Sciences, No. 509 Nanchang road, Lanzhou 730000, China}

\author{Ziqi Zhang}
\email{zhangziqi@impcas.ac.cn}
\affiliation{Institute of Modern Physics, Chinese Academy of Sciences, No. 509 Nanchang road, Lanzhou 730000, China}

\author{Kaiyu Fu} 
\email{kaiyufu94@gmail.com}
\affiliation{Institute of Modern Physics, Chinese Academy of Sciences, No. 509 Nanchang road, Lanzhou 730000, China}

\author{V.A.~Karmanov} 
\email{karmanov@sci.lebedev.ru}
\affiliation{Lebedev Physical Institute, Leninsky Prospekt 53, 119991 Moscow, Russia}

\date{\today}

\begin{abstract}
The relativistic wave function of $^3$He nucleus is calculated in the framework of Light-Front Dynamics. 
It is determined by 32 spin-isospin components, each of which depends on five scalar variables. 
For NN interaction, the one-boson exchange model is assumed, but without a potential approximation. 
The relativistic effects manifest themselves in deviation of the relativistic components from the non-relativistic input, in the appearance of the components absent in the non-relativistic limit, and in dependence of solutions on specific variables that don't exist in the non-relativistic wave function.
\end{abstract}

\maketitle

\section{Introduction}\label{intro}

Nuclei are mainly treated as many-body non-relativistic systems. 
Therefore, the calculation of their wave functions and description of the nuclear properties are based on the many-body Schr\"odinger equation. 
However, this approach cannot describe the high-momentum tails of the nuclear wave function, where the nucleon momenta are relativistic and comparable to their mass.
The relativistic domain is beyond the applicability of the  Schr\"odinger equation. 
Therefore, it is of great interest since it is determined by the not yet completely established nuclear dynamics, not reducible to the potential interaction,  which in turn determines the nuclear electromagnetic form factors at large momentum transfer.

Among the approaches used to describe nuclei in the domain of relativistic nucleons, the light-front (LF) dynamics (LFD)~\cite{Dirac} (see for review~\cite{Coester92,Brodsky98}) is widely recognized as one of the most adequate. 
However, the benefits of LFD in describing the relativistic few-body bound systems, such as baryons in the quark model and light nuclei at relativistic nucleon momenta, are far from being fully exploited. 
Often, rather simple-minded models for the wave functions, which oversimplify their rich spin structure and dependence on the variables specific to LFD, are applied to the description of relativistic systems. 
This was quite justified and understandable when the trial, pioneering steps were being taken in applications of LFD to nucleon~\cite{chung}, but it is not satisfactory at present, in the era of widespread adoption of LFD and greatly increased (super)computer power.

The technical difficulties in dealing with a relativistic wave function  are caused by the increase in the number of  spin components that the relativistic wave function contains, as well as the increase in the number of kinematic variables on which each spin component depends. On the other hand, this not only increases the difficulties but also expands the range of phenomena that arise from the rich and non-trivial structure of the light-front wave function. 

In the case of the deuteron, this increase is not yet very significant. As is well known, the non-relativistic deuteron wave function is determined by two spin components, corresponding to the angular momenta $L=0$ and 2 (the usual S- and D-waves), each depending on one variable, \ie, the relative proton-neutron momentum.
While the relativistic deuteron LF wave function is determined by six spin scalar components, each of them depends on two variables 
$|\vec{k}_{\perp}|$ and $x$, instead of one relative momentum. 
In the non-relativistic limit, only two components dominate, which turn into the usual deuteron S- and D-waves depending on the combination of 
$|\vec{k}_{\perp}|$ and $x$ forming the relative momentum $q$. The LF deuteron wave function was calculated~\cite{ck95} in the framework of the explicitly covariant version of LFD~\cite{karm76}.

This explicitly covariant version (see for review~\cite{cdkm}) is a generalization of ordinary LFD, providing the convenience for its practical use.
In the ordinary LFD, the $z$ axis in the LF equation $t+z=0$ is distinguished relative to other axes which breaks explicit relativistic covariance. 
In the explicitly covariant version, the light-front plane has a general orientation, that is, the LF equation is given by $\omega\cd x=\omega_0t-\vec{\omega}\cd\vec{r}=0$, where $\omega=(\omega_0,\vec{\omega})$ is a four-vector with $\omega^2=\omega_0^2-\vec{\omega}^2=0$. 
The ordinary LFD is obtained as a particular case corresponding to $\omega=(1,0,0,-1)$. 
The on-shell amplitudes, incorporating all the contributions in a given order of the perturbation theory, don't depend on $\omega$ and coincide with those found using the Feynman rules. 
This dependence appears because of approximations. 
Off-shell amplitudes, even exact ones, depend on $\omega$. 
Wave functions, being off-shell objects, also depend on $\omega$ as a four-vector parameter. 
This dependence appears on the equal footing with dependence on the four-momenta.
Therefore, the four-vector $\omega$ participates in the construction of spin and angular momentum of a bound state wave function together with the particle momenta. 
This considerably simplifies the construction of the wave function with definite total spin. 
Though in any version of LFD, the number of the spin components naturally increases relative to the non-relativistic wave function.

In Ref.~\cite{ck95}, the relativistic deuteron LF wave function was calculated using the one-boson exchange interaction kernel.
Then this wave function was successfully used to predict the deuteron electromagnetic form factors~\cite{ck99}. 
This prediction was later well confirmed by the experimental data~\cite{Abbott}. The agreement between the theory and experiment reaffirmed the fruitfulness of explicitly covariant LFD in application to the deuteron and its electromagnetic form factors.

The next natural step would be the application of this approach to more complicated nuclei, in particular, $^3$He.
The general spin structure of the three-fermion LF wave function was established long ago~\cite{karmNWF}.
However, it turned out that the addition of just one extra nucleon, relative to deuteron, increased the technical difficulties enormously: for each isospin of the nucleon pair $t=0$ and $t=1$, the three-fermion LF wave function is determined by 16 spin components (32 spin-isospin components in total), each of them depends on five scalar variables \cite{karmNWF}, so that a couple of decades ago the technical difficulties, caused by this complexity, were almost insurmountable.

Now, thanks to the development of  powerful computer facilities, this problem has become solvable. 
In this paper, we present a solution for the relativistic $^3$He LF wave function. 
When it comes to accounting for spins, we will follow the traditional approach adopted in nuclear physics. 
Namely, we will decompose the wave function in the spin-dependent covariant basis, constructed using the Dirac spinors. 
The invariant coefficients multiplying the basis functions (16 invariant amplitudes for each pair isospin $t=0,1$, depending, as mentioned, on five variables) are obtained by solving a system of equations. 
We  will use two bases related to each other: one  simplifies the system of equations and its solution, the other is convenient for comparison with the known non-relativistic wave functions. 
Our main strategy is rather transparent and straightforward, however, the constructions of the bases, the relation between them, and the derivation of the system of equations for the corresponding coefficients are rather technical and cumbersome. 
Therefore, in the main text of the paper, we will present the principal ``milestones'' of this construction, deferring the technical details to the appendices.
Preliminary short version of this work containing part of these results was published in \cite{NTSE24}.

Another approach - Basis Light-Front Quantization - does not construct the covariant spin structures, but decomposes the wave function, in the longitudinal direction, in terms of the plane wave basis, and in the transverse direction, via the 2D oscillator basis~\cite{Vary:2009gt}. 
Dependence not only on spins but also on the longitudinal and transverse momenta is contained in the basis functions. 
Interesting results were obtained in the application of this approach, in particular, to the nucleon \cite{BLFQ}.

The plan is as follows.
In Sec.~\ref{sec2}, we discuss the general properties and  spin structure of the three-fermion wave function. 
In Sec.~\ref{basisV}, we construct the spin basis  convenient for solving the system of equations for its spin components. 
The isospin basis functions, which are the same as those in the non-relativistic approach, are also constructed in this section.
In Sec.~\ref{new}, we construct the relativistic spin basis that coincides, in the non-relativistic limit, with the basis used to obtain solution of the Schr\"odinger equation for the $^3$He non-relativistic wave function.
The relation between the components obtained in both bases is also presented. 
The system of equations for the spin components of $^3$He nucleus corresponding to the first basis in is derived in Sec.~\ref{3beq}. 
Numerical results are presented in Sec.~\ref{num}. 
Section~\ref{concl} contains the concluding remarks. 
Technical details are provided in the Appendices~\ref{appendix1}-\ref{kinemat}. 

\section{Spin structure of the three-fermion wave function in LFD}\label{sec2}

The $^3$He nucleus has quantum numbers $I(J^P) = \frac{1}{2}(\frac{1}{2}^+)$.
To construct the $^3$He wave function, we employ the spin-isospin formalism. 
However, to simplify notations, we temporarily omit the isospin degrees of freedom when discussing the three-fermion wave functions and restore them later.

In the explicitly covariant LFD, the three-body wave function depends on the four-momenta of the constituents $k_{1,2,3}$:
\begin{equation}
  \Phi=\Phi_{\sigma_1\sigma_2\sigma_3}^{\sigma}(k_1,k_2,k_3;p,\omega\tau). \label{Psi}
\end{equation}
It is an on-mass-shell amplitude since all the four-momenta are on the corresponding mass shells, $k_1^2=k_2^2=k_3^2=m^2$, $p^2=M^2$, and 
$(\omega\tau)^2=0$.
However, it is off-energy shell since the particle four-momenta, $k_1$, $k_2$, $k_3$, and $p$, don't satisfy the conservation law $k_1+k_2+k_3=p$ valid for free particles. 
The conservation law contains extra four-momentum $\omega\tau$~\cite{cdkm}:
\begin{equation}
  k_1+k_2+k_3=p+\omega\tau.\label{conserv}
\end{equation}
In the ordinary version of LFD with $\omega=(\omega_0,\omega_x,\omega_y,\omega_z)=(1,0,0,-1) \to (\omega_-,\omega_+, \omega_x,\omega_y)=(2,0,0,0)$, where $\omega_\pm=\omega_0\pm\omega_z$, the equality (\ref{conserv}) corresponds to the usual conservation of the ``plus''- and perp-components $k_{1+}+k_{2+}+k_{3+}=p_+$ and $\vec{k}_{1\perp}+\vec{k}_{2\perp}+\vec{k}_{3\perp}=\vec{p}_{\perp}$, and the non-conservation of the ``minus''-components: $k_{1-}+k_{2-}+k_{3-}=p_- + 2\tau\neq p_-$. 
$\tau$ is the measure of this non-conservation, \ie, a measure of the deviation of the on-mass-shell amplitude (wave function) from the energy shell, more specifically, the deviation of the particle's c.m.~energy from their bound state mass $M$. 
Indeed, from Eq.~(\ref{conserv}) it follows: $\tau=({\cal M}^2-M^2)/(2\omega\cd p)$, where  ${\cal M}^2=(k_1+k_2+k_3)^2$ is invariant three-body energy squared. 
These relations can be simply translated into the trivial property of any non-relativistic system: the sum of kinetic energies and masses of constituents (in the c.m.~rest frame) is always bigger than the bound state mass.

As is well known, in relativistic theory, a moving particle has no definite projection of its spin on an arbitrary direction in space (taken, \eg, for the $z$-axis).
It has a definite projection either on the direction of its momentum (helicity) or on the $z$-axis in its rest frame (eigenvalue of the Pauli-Lubansky operator). 
We will use the latter ones to label the constituent spinors.
Particles composing a bound system move with different momenta. 
Therefore, $\sigma_{1,2,3}$ in Eq.~(\ref{Psi}) are projections of the spins of constituents on the $z$-axis in the rest frames (not coinciding with each other) of each constituent, and $\sigma$ is the spin projection of the bound state in its rest frame (we assume that the total spin of the system is 1/2). 
These spin projections are associated with the fermion spinors.
Since each spin projection can obtain two values: $\sigma_{1,2,3}=
\pm \frac{1}{2}$ and $\sigma =\pm \frac{1}{2}$, $\Phi_{\sigma_1\sigma_2\sigma_3}^{\sigma}$ contains $2\times 2\times 2\times 2=16$ elements.
Note that in the case of deuteron (two fermions forming a system with the total angular momentum $J=1$), the number of elements is $2\times 2\times 3=12$.
However, for the deuteron, due to  parity conservation, there are relations between these matrix elements reducing the number of independent elements by a factor of 2 -- from 12 to 6. 
This just corresponds to six spin components of the deuteron LF wave function~\cite{ck95}, mentioned above. 
However, as will be explained below, such a reduction does not happen for a three-body system.

To ensure the permutation properties,  we represent the full three-body wave function as a sum of the Faddeev components $\Phi_{12}$ with the arguments, for identical particles, permuted by the cyclic permutation:
\begin{equation}\label{Psi1}
  \Psi(1,2,3)=\Phi_{12}(1,2,3)+\Phi_{12}(2,3,1)+\Phi_{12}(3,1,2).
\end{equation}
The permutations also include spin and isospin projections.
If $\Phi_{12}(1,2,3)$ is antisymmetric under the permutation of only the first pair of arguments  $\Phi_{12}(2,1,3)=-\Phi_{12}(1,2,3)$,  the full function $\Psi(1,2,3)$ is antisymmetric under permutation in any pair. 
Just the Faddeev component $\Phi_{12}(1,2,3)$ without considering the isospin degrees of freedom will be decomposed into 16 basis functions. 
The coefficients of this decomposition will be found from the system of  integral equations. 

Important ingredients of our consideration will be the construction of two bases, used for this decomposition and both containing 16 elements. These bases are linearly related to each other but useful for different aims. 
One of the bases, denoted $V_{ij}$ with $i,j=1,2,3,4$, is constructed in the next section. 
The Faddeev components corresponding to the pair isospins $t=0,1$ of the constituents 1,2  are represented as follows:
\begin{equation}
  \Phi^{(0,1)}_{12}(1,2,3)=\sum_{i,j=1}^4g^{(0,1)}_{ij}(1,2,3)V_{ij},\label{via_gij}
\end{equation}
Each of these two sums contains 16 terms.
$g^{(0,1)}_{ij}(1,2,3)$ are the invariant functions depending on the particle momenta. 
The full  Faddeev component is a superposition of two isospin states:
\begin{equation}
  \Phi_{12}(1,2,3)=\Phi^{(0)}_{12}(1,2,3)\xi^{(0)}+\Phi^{(1)}_{12}(1,2,3)\xi^{(1)},\label{via_gij_f}
\end{equation}
where the isospin functions $\xi^{(0,1)}(1,2,3)$ corresponding to the pair isospins $t=0,1$ and to the total isospin 1/2 will be presented in Sec. \ref{isospin}.
Taking the isospin into account, we get 32 components $g^{(0,1)}_{ij}(1,2,3)$ - 16 for $t=0$ and 16 for $t=1$.

The advantage of the basis $V_{ij}$, constructed in Sec.~\ref{basisV}, lies in its simplicity and orthonormalization: 
\begin{align*}
  \frac{1}{2}V^{\dagger}_{i^\prime j^\prime }V_{ij}=\delta_{i^\prime i}\delta_{j^\prime j}.
\end{align*}
The product $ \frac{1}{2}{V'}^{\dagger}V$ implies sum over the nucleon spin projections and averaging over the $^3$He spin projection (the latter results in the factor $\frac{1}{2}$, see Eq. (\ref{normS}) below.)

The equation for the wave function $\Phi^{(0,1)}_{12}(1,2,3)$ will be reduced to the system of equations for the components $g^{(0,1)}_{ij}(1,2,3)$.

In terms of another basis constructed in Sec.~\ref{new} and denoted $\chi_n$ with $n=1,\ldots,16$,  the decomposition of $\Phi_{12}$ reads:
\begin{equation}\label{via_psin}
  \Phi_{12}^{(0,1)}(1,2,3)=\sum_{n=1}^{16}\chi_n\psi_n^{(0,1)}(1,2,3) .
\end{equation}
The basis functions $\chi_n$ are not as simple to orthogonalize as $V_{ij}$. The  orthonormalization is
\begin{align*}
  \frac{1}{2}\int \mathrm{d}\Omega_{\vec{k}}\mathrm{d}\Omega_{\vec{q}} \chi_i^\dagger\chi_j = \delta_{ij},
\end{align*}
where $\mathrm{d}\Omega = \sin\theta\mathrm{d}\varphi\mathrm{d}\theta$. $\vec{k}$ and $\vec{q}$ are the Jacobi momenta defined in Eq.~(\ref{Jacpqa}) below.

The advantage of this basis lies in its following property: in the non-relativistic limit, some functions $\psi_n$ turn into the known non-relativistic components, found from the three-body Schr\"odinger equation, whereas other components $\psi_n$ disappear (turn into zero).  
The components $g_{ij}$ and $\psi_n$ are linearly related to each other by Eq.~(\ref{relat}) below.
We will solve the equations (\ref{gij}) for the components $g^{(t)}_{ij}$, by the equation inverse to Eq.~(\ref{relat}) we will find the components $\psi_n$ and  compare them with the non-relativistic wave functions to analyze the influence of the relativistic effects.
Our main strategy is rather transparent and straightforward, however, the constructions of the bases, the relation between them, the derivation of the kernels and the system of equations for the corresponding coefficients are rather technical and cumbersome.

\section{Orthonormalized basis $V_{ij}$}\label{basisV}

There is some freedom in constructing the basis, since, instead of the initially constructed elements, we can take their linear combinations to obtain another basis.
Therefore, to simplify the equation, we will first construct the simplest and orthonormalized basis. 
It has the form:
\begin{align*}
  V_{ij}\propto [\bar{u}_{\sigma_1}(k_1)T_i U_c\bar{u}_{\sigma_2}(k_2)]\,[\bar{u}_{\sigma_3}(k_3)S_j u^{\sigma}(p)].
\end{align*}
Here $\bar{u}_{\sigma_{1,2,3}}(k_{1,2,3})$ are the conjugated nucleon spinors, $u^{\sigma}(p)$ is the spinor of the bound system (the nucleus $^3$He in our case).
$T_i$ and $S_j$ are the $4\times 4$-matrices (in the bi-spinor indexes), constructed to provide the orthogonality as clarified in Eqs.~(\ref{normS}) and (\ref{normT}) below, $U_c=\gamma_2\gamma_0$ is the charge conjugation matrix. 
It is absent in the second factor which contains $[\bar{u}(k_3)\ldots u^{\sigma}(p)]$, but it appears in the first factor to ensure the correct transformation properties, since both spinors in it are conjugated: $[\bar{u}_{\sigma_1}(k_1)\ldots\bar{u}_{\sigma_2}(k_2)]$. 
The transformation properties of $U_c\bar{u}_{\sigma_2}(k_2)$ are the same as for $u_{\sigma_2}(k_2)$ and opposite relative to the space inversion P.

In the LFD framework, we construct the following four spin structures $S_j$:
\begin{align}
  S_1(3)&=\frac{1}{N_{S_1}}\left(2x_3-(m+x_3 M)
  \displaystyle{\frac{\hat{\omega}}{\omega\cd p}}\right),
  \label{eq:wf2-1}\\
  S_2(3)&=\frac{1}{N_{S_2}}\frac{m}{\omega\cd p}\hat{\omega},
  \label{eq:wf2-2}\\
  S_3(3)&=\frac{i}{N_{S_3}}\left(2x_3
  -(m-x_3 M)\displaystyle{\frac{\hat{\omega}}{\omega\cd p}}\right)\gamma_5,
  \label{eq:wf2-3}\\
  S_4(3)&=\frac{i}{N_{S_4}}\frac{m}{\omega\cd p}\hat{\omega}\gamma_5,
  \label{eq:wf2-4}
\end{align}
where $m$ is the constituent mass, $M$ is the bound state mass and  $x_3=\frac{\omega\cd k_3}{\omega\cd p}$.  
In each $S_j(3)$ the argument $(3)$ indicates that $S_j(3)$ depends on $x_3$. 
Similar notations are adopted for other functions. 
Sandwiched with spinors, $\bar{u}_{\sigma_3}(k_3)S_{1,2}u^{\sigma}(p)$ are scalars, while $\bar{u}_{\sigma_3}(k_3)S_{3,4}u^{\sigma}(p)$ are pseudoscalars.

The structures (\ref{eq:wf2-1}-\ref{eq:wf2-4}) satisfy the following orthogonality condition:
\begin{align}
  &\frac{1}{2}\sum_{\sigma_3\sigma}[\bar{u}_{\sigma_3}(k_3)S_{j} u^{\sigma}(p)]^{\dagger}[\bar{u}_{\sigma_3}(k_3)S_{j^\prime } u^{\sigma}(p)]
  \nonumber\\
  &=\frac{1}{2}\sum_{\sigma_3\sigma} [\bar{u}_{\sigma}(p)\bar{S}_j u_{\sigma_3}(k_3)][\bar{u}_{\sigma_3}(k_3)S_{j^\prime } u^{\sigma}(p)]
  \nonumber\\
  &=\frac{1}{2}\Tr[\bar{S}_j(\hat{k}_3+m)S_{j^\prime }(\hat{p}+M)]=\delta_{jj^\prime },\label{normS}
\end{align}
where $\bar{S}_j=\gamma_0 S_{j}^{\dagger}\gamma_0$.
That is,
\begin{align}
  \bar{S}_{1,2}&=S_{1,2}, 
  \label{eq:S3bar_1}\\
  \bar{S_3}&=\frac{i}{N_{S_3}}\gamma_5\left(2x_3-(m-x_3 M)\frac{\hat{\omega}}{\omega\cd p}\right),
  \label{eq:S3bar_2}\\
  \bar{S}_4&=\frac{i}{N_{S_4}}m\gamma_5\frac{\hat{\omega}}{\omega\cd p},\label{eq:S3bar_3}
\end{align}
where the normalization factors,
\begin{equation}\label{wf3a}
  N_{S_j}=
  \left\{
  \begin{array}{ll}
  2\sqrt{x_3\vec{R}_{3\perp}^2} &\mbox{for $j=1,3$}
  \\
  2\sqrt{x_3m^2} &\mbox{for $j=2,4$}
  \end{array}
  \right.,
\end{equation} 
ensure the equation (\ref{normS}).
We use here the LF variables $\vec{R}_{i\perp}$ and $x_i$ defined in Appendix~\ref{app1}. 
In particular, $\vec{R}_{3\perp}$ is the component of the four-vector $R_3=k_3-x_3p=(R_{30},\vec{R}_{3\perp},\vec{R}_{3||})$ that is orthogonal to the spatial part of $\omega$, \ie, $\vec{\omega}$.
That is, $\vec{\omega}\cd \vec{R}_{3\perp}=0$, and $\vec{R}_{3||}||\vec{\omega}$. 
Since, by construction, $\omega\cd R_3=0$, we have $R^2_3=-\vec{R}^{\,2}_{3\perp}$.  
As mentioned, we introduce the factor $\frac{1}{2}$ in Eq.~(\ref{normS}) because when normalizing the wave function we use  
$\frac{1}{2}\sum_{\sigma_1\sigma_2\sigma_3\sigma}|\Phi^{\sigma}_{\sigma_1\sigma_2\sigma_3}|^2$ as the sum over the spin projections $\sigma_{1,2,3}$ of nucleons and average it over the projection $\sigma$ of the nucleus $^3$He.

Similarly, we construct another four spin structures $T_i$:
\begin{align}
  T_1(1,2,3)&=\frac{1}{N_{T_1}}\left(\frac{2x_1x_2}{x_1+x_2}-
  \displaystyle{\frac{m\hat{\omega}}{\omega\cd p}}
  \right)i\gamma_5,
  \label{eq:wf4_1}\\
  T_2(1,2,3)&=\frac{1}{N_{T_2}}\displaystyle{\frac{m\hat{\omega}}{\omega\cd p}}i\gamma_5,
  \\
  T_3(1,2,3)&=\frac{1}{N_{T_3}}\left(\frac{2x_1x_2}{x_1+x_2} + \frac{x_1 - x_2}{x_1 + x_2}
  \displaystyle{\frac{m\hat{\omega}}{\omega\cd p}}\right),
  \\
  T_4(1,2,3)&=\frac{1}{N_{T_4}}\displaystyle{\frac{m\hat{\omega}}{\omega\cd p}},\label{eq:wf4_4}
\end{align}
where
\begin{align*}
  x_i=\frac{\omega\cd k_i}{\omega\cd p}.
\end{align*}
The notations follow $T_i(1,2,3)\equiv T_i(x_1,x_2,x_3)$.
Sandwiched with the spinors, $\bar{u}_{\sigma_1}(k_1)T_{1,2}U_c\bar{u}_{\sigma_2}(k_2)$ are scalars, while 
$\bar{u}_{\sigma_1}(k_1)T_{3,4}U_c\bar{u}_{\sigma_2}(k_2)$ are pseudoscalars.

The structures (\ref{eq:wf4_1}-\ref{eq:wf4_4}) satisfy the following orthogonality and normalization conditions:
\begin{align}
  &\sum_{\sigma_1\sigma_2} [\bar{u}_{\sigma_1}(k_1)T_{i}U_c
  \bar{u}_{\sigma_2}(k_2)]^{\dagger}
  [\bar{u}_{\sigma_1}(k_1)T_{i^\prime }U_c\bar{u}_{\sigma_2}(k_2)]
  \nonumber\\
  &=-\sum_{\sigma_1\sigma_2} [u_{\sigma_2}(k_2)U_c \bar{T}_{i}
  u_{\sigma_1}(k_1)]
  [\bar{u}_{\sigma_1}(k_1)T_{i^\prime }U_c\bar{u}_{\sigma_2}(k_2)]
  \nonumber\\
  &=-\Tr[\bar{T}_i(\hat{k}_1+m)T_{i^\prime }(-\hat{k}_2+m)]=\delta_{ii^\prime },\label{normT}
\end{align}
where $\bar{T}_i=\gamma_0 T_i^{\dagger}\gamma_0.$
The sign ``minus'' in $(-\hat{k}_2+m)$ originates from $U_c\hat{k}^tU_c=-\hat{k}$ ($\hat{k}^t$ is the transposed matrix), and the total sign ``minus'' at 
$-\Tr[\ldots]$ goes from $\gamma^0U_c=-U_c\gamma^0$.
That is,
\begin{align}
  \bar{T}_1&=\frac{i}{N_{T_1}}\gamma_5\left[\frac{2x_1x_2}{x_1+x_2} - \frac{m\hat{\omega}}{\omega\cd p}\right],
  \label{Tbar1}\\
  \bar{T}_2&=\frac{i}{N_{T_2}}\gamma_5\frac{m\hat{\omega}}{\omega\cd p},\quad \bar{T}_{3,4}=T_{3,4}.\label{Tbar2}
\end{align}
The normalization factors $N_{T_i}$ read:
\begin{equation}
  N_{T_i}=
  \left\{
  \begin{array}{ll}
  \sqrt{8x_1x_2\vec{R}_{12\perp}^2}&\mbox{for $i=1,3$}
  \\
  \sqrt{8x_1x_2 m^2} &\mbox{for $i=2,4$}
  \end{array}
  \right.,\label{wf5a}
\end{equation} 
with
\begin{align*}
  \vec{R}_{12\perp}=\vec{R}_{1\perp}+\frac{x_1}{1-x_3}\vec{R}_{3\perp}.
\end{align*}

As noticed above, when sandwiched with spinors, the matrices $S_{1,2}$ and $T_{1,2}$ become scalars while $S_{3,4}$ and $T_{3,4}$ become pseudoscalars. As a result, the products,
\begin{equation}
 [\bar{u}_{\sigma_1}(k_1)T_iU_c\bar{u}_{\sigma_2}(k_2)]\,[\bar{u}_{\sigma_3}(k_3)S_ju^{\sigma}(p)],\label{product}
\end{equation}
contain 8 scalars and 8 pseudoscalars. 
In this situation, when describing a two-body system with the positive parity, \eg, deuterons, one takes only the scalar elements, which reduces the number of the basis elements by a factor of 2. 
However, as mentioned above, this does not happen in a relativistic three-body system.

From the momenta available in the  relativistic three-body system, we can construct another pseudoscalar
\begin{align}
  C_{\rm{ps}}(k_1,k_2)&=
  \frac{\varepsilon^{\mu\nu\rho\gamma}k_{1\mu}k_{2\nu}p_{\rho}\omega_{\gamma}}{|\varepsilon^{\mu\nu\rho\gamma}k_{1\mu}k_{2\nu}p_{\rho}\omega_{\gamma}|}
  \nonumber\\
  &=\frac{1}{\omega\cd p}\frac{\varepsilon^{\mu\nu\rho\gamma}k_{1\mu}k_{2\nu}p_{\rho}\omega_{\gamma}}
  {\sqrt{\vec{R}_{1\perp}^2\vec{R}_{2\perp}^2-(\vec{R}_{1\perp}\cd\vec{R}_{2\perp})^2}}.\label{wf6}
\end{align}
It is antisymmetric relative to the permutation: $C_{\rm{ps}}(k_1,k_2)=-C_{\rm{ps}}(k_2,k_1)$.
Its square is 1:
\begin{align*}
  C^2_{\rm{ps}}(k_1,k_2)=1.
\end{align*}
We also define the following factor:
\small
\begin{align}
  &C_{ij}=
  \left\{
  \begin{array}{ll}
  1 & \mbox{if $i=1,2, j=1,2$ or $i=3,4,j=3,4$}  
  \\
  C_{\rm{ps}}(k_1,k_2)& \mbox{if $i=1,2,j=3,4$ or $i=3,4,j=1,2$} 
  \end{array}
  \right. .\label{Cij}
\end{align}
\normalsize
Then, by multiplying the pseudoscalar products from Eq. (\ref{product}) by this factor, we construct 16 scalar basis functions:
\begin{equation}
  V_{ij}=C_{ij} [\bar{u}_{\sigma_1}(k_1)T_iU_c\bar{u}_{\sigma_2}(k_2)]\,[\bar{u}_{\sigma_3}(k_3)S_ju^{\sigma}(p)].\label{Vij}
\end{equation}
The conjugated functions read:
\begin{equation}
  V^{\dagger}_{ij}=-C_{ij} [u_{\sigma_2}(k_2)U_c\bar{T}_iu_{\sigma_1}(k_1)]\,[\bar{u}_{\sigma}(p)\bar{S}_ju^{\sigma_3}(k_3)]. \label{Vijc}
\end{equation}
The orthogonality and normalization condition has the form:
\begin{equation}
  \frac{1}{2}V^{\dagger}_{i^\prime j^\prime }V_{ij}=\delta_{ii^\prime } \delta_{jj^\prime },\label{Vijnorm}
\end{equation}
where the product implies sum over the spin projections $\sigma$ and $\sigma_{1,2,3}$ simplifying to the traces like in Eqs.~(\ref{normS}) and (\ref{normT}).

The pseudoscalar factor $C_{\rm ps}$ is important for constructing the full set of spin components with given parity.
For the non-relativistic two- and three-body wave function \cite{kolybasov}, the parity conservation reduces the number of basis functions by a factor of 2. 
As it was noticed long ago \cite{kolybasov}, this reduction does not occur in the $n$-body non-relativistic wave function for $n\geq 4$. In the case $n=4$, the momentum conservation law $\vec{k}_1+\vec{k}_2+\vec{k}_3+\vec{k}_4=0$ in the rest frame $\vec{p}=0$, leaves three independent momenta, which allow to construct the pseudoscalar $c_{\rm ps}=\vec{k}_3\cd[\vec{k}_1\times \vec{k}_2]$ (here $[\vec{k}_1\times \vec{k}_2]$ is the cross product).

This is exactly what happens in the three-body LF wave function. Although it depends only on three particle momenta, it also includes an additional vector 
$\vec{\omega}$, which allows to construct the pseudoscalar (\ref{wf6})
and 16 basis elements with the same parity for the $^3$He wave function. These 16 elements are explicitly constructed and presented in Eq.~(\ref{Vij}).
Note that the eight elements with indexes $ij=13$, 14, 23, 24, 31, 32, 43, 44 are symmetric under the permutation of the particles $12\leftrightarrow 21$, while the remaining eight elements are anti-symmetric.

\subsection{Isospin basis}\label{isospin}

Let us now construct the isospin basis functions. 
The total isospin of both $^3$He and $^3$H is 1/2. 
The Faddeev component $\Phi_{12}$ is a superposition of two states, with isospin of the nucleon pair 12 being 0 and 1. The corresponding isospin functions $\xi^{(0)}$ and $\xi^{(1)}$, which correspond to the total isospin 1/2 and the pair isospins 0 and 1, respectively, are expressed through the Clebsch-Gordan coefficients as
\begin{align}
    \xi^{(0)}(123)=
    \xi^{(0)\tau}_{\tau_1\tau_2\tau_3}&=C^{00}_{\frac{1}{2}\tau_1\,\frac{1}{2}\tau_2} 
    C^{\frac{1}{2}\tau}_{00\,\frac{1}{2}\tau_3} , 
    \label{eq:xi01a_1}\\
    \xi^{(1)}(123)=
    \xi^{(1)\tau}_{\tau_1\tau_2\tau_3}&=\sum_{\tau_{12}}C^{1\tau_{12}}_{\frac{1}{2}\tau_1\,\frac{1}{2}\tau_2}
    C^{\frac{1}{2}\tau}_{1\tau_{12}\,\frac{1}{2}\tau_3} . 
    \label{eq:xi01a_2}
\end{align}
They are orthonormalized:
\begin{align}
    \xi^{0\dagger}\xi^{0}&=\sum_{\tau_1\tau_2\tau_3}\xi^{(0)\tau}_{\tau_1\tau_2\tau_3}\xi^{(0)\tau^\prime }_{\tau_1\tau_2\tau_3}=\delta^{\tau\tau^\prime },
    \label{eq:ortho_xi1}\\
    \xi^{1\dagger}\xi^{1}&=\sum_{\tau_1\tau_2\tau_3}\xi^{(1)\tau}_{\tau_1\tau_2\tau_3}\xi^{(1)\tau^\prime }_{\tau_1\tau_2\tau_3}=\delta^{\tau\tau^\prime },
    \\
    \xi^{0\dagger}\xi^{1}&=\xi^{1\dagger}\xi^{0}=\sum_{\tau_1\tau_2\tau_3}\xi^{(0)\tau}_{\tau_1\tau_2\tau_3}\xi^{(1)\tau^\prime }_{\tau_1\tau_2\tau_3}=0.\label{eq:ortho_xi3}
\end{align}
For shortness, we omit the indexes $\tau_1$, $\tau_2$, $\tau_3$, and $\tau$ when it does not lead to misunderstanding.

The isospin functions after permutations,
\begin{align}
    \xi^{(0)}(312)&=\xi^{(0)\tau}_{\tau_3\tau_1\tau_2}=C^{00}_{\frac{1}{2}\tau_3\,\frac{1}{2}\tau_1}
    C^{\frac{1}{2}\tau}_{00\,\frac{1}{2}\tau_2}
    \label{eq:xi312_1},\\
    \xi^{(1)}(312)&=\xi^{(1)\tau}_{\tau_3\tau_1\tau_2}=\sum_{\tau_{31}}C^{1\tau_{31}}_{\frac{1}{2}\tau_3\,\frac{1}{2}\tau_1}
    C^{\frac{1}{2}\tau}_{1\tau_{31}\,\frac{1}{2}\tau_2}\label{eq:xi312_2},
\end{align}
(and similarly for $\xi^{(0,1)}(231)$) can be decomposed through the functions (\ref{eq:xi01a_1}) and (\ref{eq:xi01a_2}):
\begin{align}
  \xi^{(0)}(312)&=-\frac{1}{2}\xi^{(0)}(123)-\frac{\sqrt{3}}{2}\xi^{(1)}(123),
  \label{eq:xi_pert1}\\
  \xi^{(1)}(312)&=\frac{\sqrt{3}}{2}\xi^{(0)}(123)-\frac{1}{2}\xi^{(1)}(123),
  \\
  \xi^{(0)}(231)&=-\frac{1}{2}\xi^{(0)}(123)+\frac{\sqrt{3}}{2}\xi^{(1)}(123),
  \\
  \xi^{(1)}(231)&=-\frac{\sqrt{3}}{2}\xi^{(0)}(123)-\frac{1}{2}\xi^{(1)}(123).\label{eq:xi_pert4}
\end{align}
From these formulas, using orthogonality, we can calculate the products
\begin{align}
  &\eta(123,klm;t,t^\prime ;T)
  =\left\{
  \begin{array}{ll}
  \xi^{t\dagger}(123)\xi^{t^\prime }(klm),&T=0
  \\
  \xi^{t\dagger}(123)(\vec{\tau}_1\cd\vec{\tau}_2)\xi^{t^\prime }(klm),&T=1
  \end{array}
  \right.
  \nonumber\\
  &=\left\{
  \begin{array}{ll}
  \phantom{-3}\xi^{t\dagger}(123)\xi^{t^\prime }(klm),&T=0,
  \\
  -3\xi^{t\dagger}(123)\xi^{t^\prime }(klm),&T=1,\;t=0,
  \\
  \phantom{-3}\xi^{t\dagger}(123)\xi^{t^\prime }(klm),&T=1,\;t=1,
  \end{array}
  \right.
  \label{eta_dif}
\end{align}
which will enter as factors in the interaction kernel. The indices $klm$ run through the cyclic permutations $123$, $312$, and $231$. The index $t=0,1$ is the isospin of the pair 12, $t^\prime =0,1$ is the isospin of the pair $kl$, and $T=0,1$ is the isospin of the exchanged meson. For $T=1$, the factor $(\vec{\tau}_1\cd\vec{\tau}_2)$ appears in the definition of 
$\eta(123,klm;t,t^\prime ;T)$.  For $T=0$, this factor is replaced by 1. The values of the coefficients $\eta(123,klm;t,t^\prime ;T)$ are given in Table~\ref{tab2}.

\begin{table}
  \begin{center}
  \caption{
  The isotopic coefficients $\eta(123,klm;t,t^\prime ;T)$ defined in Eq.~(\ref{eta_dif}). $t$ and $t^\prime $ are the isospins of the nucleon pairs 12, $T$ is the isospin of the exchanged meson.}\label{tab2}
  \begin{tabular}{ccc|ccc}
    \hline \hline
    \multicolumn{3}{c|}{isospins} & \multicolumn{3}{c}{$klm$} \\ \hline
          $T$ & $t$ &  $t^\prime $ &123 &321&213      \\ \hline
          0& 0 &0&1&$-\frac{1}{2}$&$-\frac{1}{2}$\\
          0& 0 &1&$-3$&$\frac{3}{2}$&$\frac{3}{2}$\\
          0& 1&0&0&$\frac{\sqrt{3}}{2}$&$\frac{\sqrt{3}}{2}$\\
          0& 1&1&0&$-\frac{3\sqrt{3}}{2}$&$\frac{3\sqrt{3}}{2}$\\
          1& 0 &0&0&$\frac{\sqrt{3}}{2}$&$\frac{\sqrt{3}}{2}$\\
          1& 0 &1&0&$-\frac{\sqrt{3}}{2}$&$\frac{\sqrt{3}}{2}$\\
          1& 1&0&1&$-\frac{1}{2}$&$-\frac{1}{2}$\\
          1& 1&1&1&$-\frac{1}{2}$&$-\frac{1}{2}$\\ \hline  \hline
  \end{tabular}
  \end{center}
\end{table}

\subsection{Normalization integral}\label{normPhi}

The $n$-body normalization integral is given by Eq.~(3.23) from Ref.~\cite{cdkm}, which for a three-body system obtains the following form:
\begin{align}
  &I=(2\pi)^3\int\frac{1}{2}
  \sum_ {\vspace{-0.2cm}
\begin{array}{l}
          \scriptstyle{\sigma_1\sigma_2\sigma_3\sigma}
  \vspace{-0.2cm}
  \\
  \scriptstyle{ \tau_1\tau_2\tau_3\tau}
                          \end{array}}
  \left|\Psi^{\sigma;\tau}_{\sigma_1\sigma_2\sigma_3;\tau_1\tau_2\tau_3}\right|^2
  \delta^{(2)}(\sum_i^3 \vec{k}_{i\perp})
  \nonumber\\
  &\times
  \delta(\sum_i^3 x_i-1)
  2\frac{\mathrm{d}^2k_{1\perp}\mathrm{d}x_1}{(2\pi)^3 2x_1}  \frac{\mathrm{d}^2k_{2\perp}\mathrm{d}x_2}{(2\pi)^3 2x_2}  \frac{\mathrm{d}^2k_{3\perp}\mathrm{d}x_3}{(2\pi)^3 2x_3}.
  \label{norm}
\end{align}
The factor $\frac{1}{2}$ arises from averaging over the spin projection $\sigma$ of the nucleus.
The normalization condition requires $I=1$. 

The function $\Psi^{\sigma}_{\sigma_1\sigma_2\sigma_3}$ is given by the sum (\ref{Psi1}). 
Its square consists of the square 
of\footnote{To simplify the notations, we omit sometimes the pair isospin index $t$, when this does not lead to misunderstanding.} $\Phi_{12}(1,2,3)$, the other squares of two cyclic permutations, and their products, such as $2\Phi^{\dagger}_{12}(2,3,1)\Phi_{12}(1,2,3)$.
Because of the identities of the particles, the integrals for the terms obtained by the cyclic permutations are the same. 
As a result, being expressed via Faddeev components, the normalization integral (\ref{norm}) reduces to 
\begin{align}
  I&=3(2\pi)^3\int\frac{1}{2} \sum_{\sigma_1\sigma_2\sigma_3\sigma;\tau_1\tau_2\tau_3\tau}\Bigl[|\Phi_{12}(1,2,3)|^2
  \nonumber\\
  &+2\Phi^{\dagger}_{12}(2,3,1)\Phi^{}_{12}(1,2,3)\Bigr]D,\label{norm1}
\end{align}
where $D$ denotes the integration volume in (\ref{norm}), including the delta-functions.

The first term in Eq.~(\ref{norm1}) contains the product $V^{\dagger}_{i^\prime j^\prime }(1,2,3)V_{ij}(1,2,3)$ which is eliminated due to the orthonormalization (\ref{Vijnorm}). The analogous product in the second term is reduced to the trace:
\begin{align}
  &T_{i^\prime j^\prime ij}=V^{\dagger}_{i^\prime j^\prime }(2,3,1)V_{ij}(1,2,3)=
  \Tr\Bigl[\bar{T}_{i^\prime }(1,2,3)(\hat{k}_1+m)
  \nonumber\\
  &\times S_j(1)(\hat{p}+M)\bar{S}_{j^\prime }(3)(\hat{k}_3+m)\tilde{T}_i(2,3,1)(-\hat{k}_2+m)\Bigr],\label{Tijij}
\end{align}
where $\tilde{T}_i(2,3,1)=U_cT^t_i(2,3,1)U_c.$

Due to orthonormalization of the isospin functions, Eqs.~(\ref{eq:ortho_xi1}-\ref{eq:ortho_xi3}), the first line in Eq.~(\ref{norm1}) is reduced to
\begin{align*}
  g_{ij}^{(0)}(1,2,3)g_{i^\prime j^\prime }^{(0)}(1,2,3)+g_{ij}^{(1)}(1,2,3)g_{i^\prime j^\prime }^{(1)}(1,2,3).
\end{align*}
The second term in Eq.~(\ref{norm1}) contains:
\begin{align}
  &G_{i^\prime j^\prime ij}\equiv
  -\frac{1}{2}\Bigl[g^{(0)}_{i^\prime j^\prime }(2,3,1)g^{(0)}_{ij}(1,2,3)
   \nonumber\\ 
  &
  +g^{(1)}_{i^\prime j^\prime }(2,3,1)g^{(1)}_{ij}(1,2,3)\Bigr]
  +\frac{\sqrt{3}}{2}\Bigl[g^{(0)}_{i^\prime j^\prime }(2,3,1)g^{(1)}_{ij}(1,2,3)
  \nonumber\\ 
  &
  -g^{(1)}_{i^\prime j^\prime }(2,3,1)g^{(0)}_{ij}(1,2,3)\Bigr].\label{Gijij}
\end{align}
The coefficients here are determined by Eqs.~(\ref{eq:xi_pert1}-\ref{eq:xi_pert4}). 
Specifically, the first coefficient is given by the product 
$\xi^{(0)\dagger}(1,2,3)\xi^{(0)}(2,3,1)=-\frac{1}{2}$, as follows from Eqs.~(\ref{eq:xi_pert1}-\ref{eq:xi_pert4}), etc.
The products of the remaining  isotopic functions are calculated similarly. 
As a result, we finally obtain:
\begin{align}
  I&=I^{(0)}+I^{(1)}+I^{}_{\mathrm{mix}},\label{norm2}
\end{align}
where 
\begin{align}
  I^{(t)}&=\frac{3}{(2\pi)^3}\int 
  \sum_{ij}\Bigl[g_{ij}^{(t)}(1,2,3)\Bigr]^2D,\, \mathrm{for}\, t=0,1,\label{norm2_2}\\
  I^{}_{\mathrm{mix}}&=\frac{3}{(2\pi)^3}\int 
  \sum_{ij}\sum_{i^\prime j^\prime ij}T_{i^\prime j^\prime ij}G_{i^\prime j^\prime ij}D. \label{norm2_3}
\end{align}
Here, $T_{i^\prime j^\prime ij}$ and $G_{i^\prime j^\prime ij}$ are defined in Eqs.~(\ref{Tijij}) and (\ref{Gijij}), respectively.

\section{Another spin basis}\label{new}
The advantage of the basis constructed in the previous section lies in its simplicity and orthonormalization, which significantly simplifies both the equation for the wave function and the form of the kernel. 
However, its disadvantage is that it does not turn into a basis for constructing non-relativistic wave functions in the non-relativistic limit.
Therefore, it is not possible to directly compare  relativistic spin components found in the basis $V_{ij}$ with the non-relativistic ones found by solving the Schr\"odinger equation. 

In this section, we will construct another LF relativistic basis   whose components, in the non-relativistic limit, are reduced to those of  the non-relativistic basis. This construction enables us to compare the relativistic solutions with the non-relativistic ones, allowing us to assess the influence of the relativistic effects.

We begin by describing the non-relativistic basis employed in Ref.~\cite{baru} for calculating  the non-relativistic $^3$He wave function. This basis will then be generalized to the relativistic framework.

\subsection{Non-relativistic basis}\label{nrb}

\begin{table}
  \begin{center}
    \caption{Spins, angular momenta and isospins forming the Faddeev component  (\ref{Phi123}).
    $s_{12}$ is the total spin of the particles 12. 
    $l_{12}$ is their angular momentum. 
    $j_{12}$ is total two-body angular momentum, formed by $s_{12}$ and $l_{12}$.
    $s_3=1/2$ is the spin of the 3rd particle. 
    $l_3$ is its angular momentum. 
    $j_3$ is the total angular momentum, formed by $s_3$ and $l_3$. 
    The total angular momenta $j_{12}$ and $j_3$ provide the total spin $J=1/2$ of $^3$He.
    $t_{12}=0,1$ is the total isospin of the particles 12. 
    $t_3=1/2$ is the isospin of the 3rd particle. 
    The pair isospin $t_{12}$ and $t_3=1/2$ provide the total isospin $t=1/2$.}\label{tab1}
    \begin{tabular}{c|ccccccc|ccc}
      \hline \hline
      & \multicolumn{7}{c|}{spins-angular momenta} & \multicolumn{3}{c}{isospins} \\ \hline
      $n$&   $s_{12}$ & $l_{12}$ &  $j_{12}$ & $s_3$ & $l_3$ & $j_3$& $J$  &   $t_{12}$ &$t_3$&$t$    \\ \hline
      1& 0 &0 & 0 & 1/2 &0& 1/2 &1/2&1&1/2&1/2\\
      2& 1 &0 & 1 & 1/2 &0& 1/2 &1/2&0&1/2&1/2\\
      3& 1 &2 & 1 & 1/2 &0& 1/2 &1/2&0&1/2&1/2\\
      4& 1 &0 & 1 & 1/2 &2& 3/2 &1/2&0&1/2&1/2\\
      5& 1 &2 & 1 & 1/2 &2& 3/2 &1/2&0&1/2&1/2\\ \hline\hline
     \end{tabular}
  \end{center}
\end{table}

The total angular momentum $J=1/2$ and total isospin $t=1/2$ are formed from the spins, angular momenta, and isospins of nucleons, as shown in Table \ref{tab1}.
The non-relativistic Faddeev  component $\Phi_{12}(1,2,3)$ is expressed as a superposition of five spin-isospin basis functions (incorporating the dominant partial waves):
\begin{equation}\label{Phi123}
\Phi_{12}(1,2,3)=\phi_1+\phi_2+\phi_3+\phi_4+\phi_5,
\end{equation}
where
\begin{align*}
  \phi_1=\tilde{\chi}_1\;\xi^{(1)}\psi_1,\; \phi_{2-5}=\tilde{\chi}_{2-5}\;\xi^{(0)}\psi_{2-5}.  
\end{align*}
$\tilde{\chi}_{1-5}$ are the spin and angular-momentum functions\footnote{We use ``tilde'' to denote the non-relativistic functions, to distinguish them from the relativistic ones  defined below.}:
\begin{align} 
  \tilde{\chi}_1&=C^{00}_{\frac{1}{2}\sigma_1\frac{1}{2}\sigma_2}
  C^{\frac{1}{2}\sigma}_{00\frac{1}{2}\sigma_3}Y_{00}Y_{00}
  =\frac{i}{4\pi\sqrt{2}}(\sigma^y)_{\sigma_1\sigma_2}\delta_{\sigma_3\sigma},\label{chinr_1}\\
  \tilde{\chi}_2&=\sum_{\sigma_{12}} C^{1\sigma_{12}}_{\frac{1}{2}\sigma_1\frac{1}{2}\sigma_2}
  C^{\frac{1}{2}\sigma}_{1\sigma_{12}\frac{1}{2}\sigma_3}Y_{00}Y_{00}
  \nonumber\\
  & =
  -\frac{i}{4\pi\sqrt{6}}\sum_i(\sigma^i\sigma^y)_{\sigma_1\sigma_2}
  (\sigma^i)_{\sigma_3\sigma},\\
  \tilde{\chi}_3&=\sum_{\sigma_{12}\sigma^\prime m}
  C^{1\sigma_{12}}_{\frac{1}{2}\sigma_1\,\frac{1}{2}\sigma_2}
  C^{1\sigma^\prime }_{2m\,1\sigma_{12}}Y_{2m}\left(\frac{\vec{k}}{|\vec{k}|}\right)
  C^{\frac{1}{2}\sigma}_{1\sigma^\prime \,\frac{1}{2}\sigma_3}Y_{00}
  \nonumber\\
  &=
  \frac{i\sqrt{3}}{8\pi}\sum_{ij}(\sigma^i\sigma^y)_{\sigma_1\sigma_2}
  (\sigma^j)_{\sigma_3\sigma}\left(\frac{k_i k_j}{\vec{k}^2}-\frac{1}{3}\delta_{ij}\right)
  ,\\
  \tilde{\chi}_4&=\sum_{\sigma_{12}\sigma^\prime m}
  C^{1\sigma_{12}}_{\frac{1}{2}\sigma_1\,\frac{1}{2}\sigma_2}
  C^{\frac{3}{2}\sigma^\prime }_{2m\,\frac{1}{2}\sigma_3}Y_{2m}\left(\frac{\vec{q}}{|\vec{q}|}\right)
  C^{\frac{1}{2}\sigma}_{1\sigma_{12}\,\frac{3}{2}\sigma^\prime }Y_{00}
  \nonumber\\
  & =
  \frac{i\sqrt{3}}{8\pi}\sum_{ij}(\sigma^i\sigma^y)_{\sigma_1\sigma_2}
  (\sigma^j)_{\sigma_3\sigma}\left(\frac{q_i q_j}{\vec{q}^2}-\frac{1}{3}\delta_{ij}\right),\\
  \tilde{\chi}_5&=\sum_{\sigma_{12}\sigma^\prime\sigma^{\prime\prime} m m^\prime}
  C^{1\sigma_{12}}_{\frac{1}{2}\sigma_1\,\frac{1}{2}\sigma_2}
  C^{1\sigma^\prime }_{2m\,1\sigma_{12}}Y_{2m}\left(\frac{\vec{k}}{|\vec{k}|}\right)
  \nonumber\\
  &\times C^{\frac{3}{2}\sigma^{\prime\prime}}_{2m^\prime \,\frac{1}{2}\sigma_3}Y_{2m^\prime }\left(\frac{\vec{q}}{|\vec{q}|}\right)
  C^{\frac{1}{2}\sigma}_{1\sigma^\prime \,\frac{3}{2}\sigma^{\prime\prime}}
  \nonumber\\
  &=-\frac{3i\sqrt{6}}{16\pi}\sum_{ij^\prime j}(\sigma^i\sigma^y)_{\sigma_1\sigma_2}
  \left(\frac{k_i k_{j^\prime }}{\vec{k}^2}-\frac{1}{3}\delta_{ij^\prime }\right)
  \nonumber\\
  &\times\left(\frac{q_{j^\prime } q_{j}}{\vec{q}^2}-\frac{1}{3}\delta_{j^\prime j}\right)
  (\sigma^j)_{\sigma_3\sigma},\label{chinr_5}
\end{align}
\normalsize
where $\sigma^i$ and $\sigma^3$ are the Pauli matrices\footnote{The relations between the middle and r.h.-sides of these equalities are obtained assuming the standard definition of the spherical function $Y_{lm}$, without the inclusion of the  extra factor $i^l$.}.
$\vec{k}$ and $\vec{q}$ are the Jacobi momenta:
\begin{equation}
  \vec{k}=\frac{1}{2}(\vec{k}_1-\vec{k}_2),\quad
  \vec{q}=\frac{1}{3}\left(\frac{\vec{k}_1+\vec{k}_2}{2}-\vec{k}_3\right),\label{Jacpq}
\end{equation}
where the momenta $\vec{k}_{1,2,3}$ are the particle momenta. 

The three-body isospin functions $\xi^{(0)}$ and $\xi^{(1)}$ are defined in Eqs.~(\ref{eq:xi01a_1}-\ref{eq:xi01a_2}).
The functions $\psi_{1-5}=\psi_{1-5}(k,q)$ are the solutions of the three-body Schr\"odinger equation, and are parametrized in Ref.~\cite{baru,green}. In this parametrization, they depend on the modules $k,q$ of the Jacobi momenta (\ref{Jacpq}).

\subsection{Transformation properties of relativistic wave functions}

In the non-relativistic system, the values $\sigma_{1,2,3}$ and $\sigma$ in Eqs.~(\ref{chinr_1}-\ref{chinr_5}) are the spin projections of  the nucleons and of the nucleus $^3$He on the $z$-axis. 
As explained below Eq.~(\ref{Psi}), $\sigma_{1,2,3}$ and $\sigma$ at $\Phi_{\sigma_1\sigma_2\sigma_3}^{\sigma}$ are the spin projections of the particles on the $z$-axis in the rest frames of each particle. 
These rest frames are different for different moving particles. 
The rotation and Lorentz transformation $g$ of the reference frame $x\to x^\prime =gx$ change the orientations of the particle spins and their projections.
Namely:
\begin{align}
  &\Phi^{\frac{1}{2}\sigma}_{\frac{1}{2}\sigma_1 \frac{1}{2}\sigma_2 \frac{1}{2}\sigma_3}(gk_1,gk_2,gp,g\omega\tau)
  =\sum_{\sigma^\prime \sigma^\prime _1\sigma^\prime _2\sigma^\prime _3}D^{(\frac{1}{2})*}_{\sigma\sigma^\prime }\{R(g,p)\}
  \nonumber\\
  &\times
  D^{(\frac{1}{2})}_{\sigma_1\sigma^\prime _1}\{R(g,k_1)\}
  D^{(\frac{1}{2})}_{\sigma_2\sigma^\prime _2}\{R(g,k_2)\}
  D^{(\frac{1}{2})}_{\sigma_3\sigma^\prime _3}\{R(g,k_3)\}
  \nonumber \\
  &\times\Phi^{\frac{1}{2}\sigma^\prime }_{\frac{1}{2}\sigma^\prime _1 \frac{1}{2}\sigma^\prime _2 \frac{1}{2}\sigma^\prime _3}(k_1,k_2,p,\omega\tau),\label{wfp6}
\end{align}
where $R(g,p)$ is the following rotation operator (Wigner rotation):
\begin{equation}\label{R}
  R(g,p)=L^{-1}(gp)gL(p),
\end{equation}
and analogously for $R(g,k_{1,2,3})$. 
These rotations are just taken into account by the rotation matrices $D^{(\frac{1}{2})}\{R(g,p)\}$ and $D^{(\frac{1}{2})}\{R(g,k_{1,2,3})\}$. 
The notations are symbolic, implying that $D^{(\frac{1}{2})}_{\sigma\sigma^\prime }$ depends on the Euler angles corresponding to the rotations $R(g,p)$ and $R(g,k_{1,2,3})$. 
These angles are not given here, we don't need their explicit expressions.

In Eq.~(\ref{wfp6}), in contrast to the transformation of the non-relativistic wave function, as mentioned above, the spin projections defined in the rest frame of each particle (which are different) are transformed by different rotation operators depending on the particle momenta. 
To represent the LF wave function in the form similar to the non-relativistic one, we, at first, introduce the representation~\cite{karm1979} 
where  {\it one and the same} rotation is applied to any spin projection, similar to the non-relativistic case.
Although this is explained in detail in Ref.~\cite{cdkm}, we provide a brief description here for completeness and clarity. 
Namely, instead of  using $\Phi$, we introduce a new function $\Psi$ related to $\Phi$ as follows:
\begin{align}
  &\Psi^{\frac{1}{2}\sigma}_{\frac{1}{2}\sigma_1 \frac{1}{2}\sigma_2 \frac{1}{2}\sigma_3}(k_1,k_2,p,\omega\tau)
  \nonumber\\
  &=\sum_{\sigma^\prime \sigma^\prime _1\sigma^\prime _2\sigma^\prime _3}
  D^{(\frac{1}{2})*}_{\sigma\sigma^\prime }\{R(L^{-1}({\cal P}),p)\}
  D^{(\frac{1}{2})}_{\sigma_1\sigma^\prime _1}\{R(L^{-1}({\cal P}),k_1)\}
  \nonumber\\
  &\times
  D^{(\frac{1}{2})}_{\sigma_2\sigma^\prime _2}\{R(L^{-1}({\cal P}),k_2)\}
  D^{(\frac{1}{2})}_{\sigma_3\sigma^\prime _3}\{R(L^{-1}({\cal P}),k_3)\}
  \nonumber \\
  &\times\Phi^{\frac{1}{2}\sigma^\prime }_{\frac{1}{2}\sigma^\prime _1 \frac{1}{2}\sigma^\prime _2 \frac{1}{2}\sigma^\prime _3}(k_1,k_2,p,\omega\tau),\label{wfp7}
\end{align}
where 
\begin{equation}\label{calP}
  {\cal P}=k_1+k_2+k_3=p+\omega\tau,
\end{equation}
and, for example, $R(L^{-1}({\cal P}),p)$ is defined by Eq.~(\ref{R}) with $g=L^{-1}({\cal P})$.   One can show~\cite{karm1979,cdkm} that  the wave function
$\Psi^{\frac{1}{2}\sigma}_{\frac{1}{2}\sigma_1 \frac{1}{2}\sigma_2 \frac{1}{2}\sigma_3}$
transforms similarly to Eq.~(\ref{wfp6}), but  with the arguments of the $D$-functions replaced by {\it one and the same} rotation operator $R(g,{\cal P})$.
Namely, under the rotations and Lorentz transformations $g$ of the reference frame $x\to x^\prime =gx$, the wave function $\Psi$ transforms as:
\begin{align}
  &\Psi^{\frac{1}{2}\sigma}_{\frac{1}{2}\sigma_1 \frac{1}{2}\sigma_2 \frac{1}{2}\sigma_3}(gk_1,gk_2,gp,g\omega\tau) =\sum_{\sigma^\prime \sigma^\prime _1\sigma^\prime _2\sigma^\prime _3}D^{(\frac{1}{2})*}_{\sigma\sigma^\prime }\{R(g,{\cal P})\}
  \nonumber\\
  &\times
  D^{(\frac{1}{2})}_{\sigma_1\sigma^\prime _1}\{R(g,{\cal P})\}
  D^{(\frac{1}{2})}_{\sigma_2\sigma^\prime _2}\{R(g,{\cal P})\}
  D^{(\frac{1}{2})}_{\sigma_3\sigma^\prime _3}\{R(g,{\cal P})\}
  \nonumber \\
  &\times\Psi^{\frac{1}{2}\sigma^\prime }_{\frac{1}{2}\sigma^\prime _1 \frac{1}{2}\sigma^\prime _2 \frac{1}{2}\sigma^\prime _3}(k_1,k_2,p,\omega\tau)\label{wfp8}.
\end{align}

We also define the following three-momenta
\begin{align}
  \vec{q}_i= L^{-1}({\cal P})\vec{k}_i = \vec{k}_i -
  \frac{\vec{\cal P}}{\sqrt{{\cal P}^2}}\left[k_{i0}
  - \frac{\vec{k}_i\cd\vec{{\cal P}}}{\sqrt{{\cal P}^2}+{\cal P}_0}\right], \label{vqi}\\
  \vec{n} = L^{-1}({\cal P})\vec{\omega}/|L^{-1}({\cal P})
  \vec{\omega}| = \sqrt{{\cal P}^2} L^{-1}({\cal P})
  \vec{\omega}/\omega\cd p, \label{vn}
\end{align}
where $ \vec{n}$ is the unit vector.
Under the rotation or Lorentz transformation $g$ over the four-vectors used for constructing these three-vectors, $\vec{q}_i$ and $\vec{n}$ are rotated by the action of the same rotation operator $R(g,{\cal P})$ which transforms in Eq.~(\ref{wfp8}) the function 
 $\Psi^{\frac{1}{2}\sigma}_{\frac{1}{2}\sigma_1 \frac{1}{2}\sigma_2 \frac{1}{2}\sigma_3}$. 
Therefore, the construction of the relativistic basis is closely parallel to the non-relativistic case, with the only essential difference being the presence of an additional three-vector $\vec{n}$.   
As a result, in the non-relativistic limit -- where the relativistic effects and dependence on $\vec{n}$ vanish -- this relativistic basis naturally reduces to the non-relativistic one.

According to the definition (\ref{vqi}), in any system of reference:
\begin{align}
  \vec{q}_1+\vec{q}_2+\vec{q}_3=L^{-1}({\cal P})(\vec{k}_1+\vec{k}_2+\vec{k}_3)=L^{-1}({\cal P})\vec{\cal P}=0.\label{sumqi}
\end{align}
Because of the conservation law (\ref{sumqi}), there are only two of the three independent vectors among $\vec{q}_{1,2,3}$. 
Consequently, as in the non-relativistic system, it is convenient to introduce two Jacobi momenta:
\begin{align}
  \vec{k}=\frac{1}{2}(\vec{q}_1-\vec{q}_2),\quad
  \vec{q}=\frac{1}{3}\left(\frac{\vec{q}_1+\vec{q}_2}{2}-\vec{q}_3\right).\label{Jacpqa}
\end{align}
These expressions coincide with Eqs.~(\ref{Jacpq}) under the replacement $\vec{k}_i\to \vec{q}_i$.  Below we will use the definitions (\ref{Jacpqa}), but we will keep for them the same notations $\vec{k}$ and $\vec{q}$ as for (\ref{Jacpq}).

Apart from the spherical functions, the non-relativistic functions $\tilde{\chi}_{1-5}$, Eqs.~(\ref{chinr_1}-\ref{chinr_5}), are constructed from the Clebsch-Gordan coefficients, or, equivalently, the Pauli matrices, and the non-relativistic spinors. These spinors transform under rotations via {\it one and the same} rotation matrix. 
For example, $\tilde{\chi}$ contains
\begin{align*}
  (\sigma^j)_{\sigma_3\sigma} p_j= w^{\dagger}_{\sigma_3}(\vec{\sigma}\cd \vec{p})w_{\sigma}
\end{align*}
with the spinors $w_{\sigma},w^{\dagger}_{\sigma_3}$.

To construct the basis for decomposing the relativistic wave function $\Psi$, Eq.~(\ref{wfp7}), and possessing the same transformation property as given in Eq.~(\ref{wfp8}), we introduce, in addition to the momenta $\vec{q}_i,\vec{n}$, Eqs.~(\ref{vqi}) and (\ref{vn}), a set of bi-spinors that transform in the same way as these three-vectors and $\Psi$.
Specifically, instead of the bi-spinor $u^{\sigma_1}(k_1)$, we define a modified bi-spinor $u_{\cal P}^{\sigma_1}(k_1)$ as,
\begin{align}
  u_{\cal P}^{\sigma_1}(k_1)=\sum_{\sigma^\prime _1}D^{(\frac{1}{2})}_{\sigma_1\sigma^\prime _1}\{R(L^{-1}({\cal P}),k_1)\}
  u^{\sigma^\prime _1}(k_1).\label{chi3}
\end{align}
Under rotations and Lorentz transformations $g$, the bi-spinor $u_{\cal P}^{\sigma_1}(k_1)$  transforms as
\begin{align}
  u^{\prime \sigma_1}_{\cal P}(gk_1)=\sum_{\sigma^\prime _1}D^{(\frac{1}{2})}_{\sigma_1\sigma^\prime _1}\{R(g,{\cal P})\}u_{\cal P}^{\sigma^\prime _1}(k_1),\label{uPp}
\end{align}
that is, by the same matrix   $D^{(\frac{1}{2})}_{\sigma_1\sigma^\prime _1}\{R(g,{\cal P})\}$, as enters in Eq.~(\ref{wfp8}).

Then we introduce the ``relativistic generalization'' of the Pauli matrices. Namely:
\begin{align}
  \vec{\sigma}_{12}^{\sigma_1\sigma_2}&=-c_1c_2\bar{u}_{\cal P}^{\sigma_1}(k_1)\Pi_+ \vec{\gamma}\Pi_- U_c \bar{u}_{\cal P}^{\sigma_2}(k_2),\label{sigmas1}
  \\
  \vec{\sigma}_{3N}^{\sigma_3\sigma}&=-c_3 c_N\bar{u}_{\cal P}^{\sigma_3}(k_3)\Pi_+ \vec{\gamma}\gamma_5\Pi_+ u_{\cal P}^{\sigma}(p),\label{sigmas2}
\end{align}
where $\vec{\gamma}$ is the spatial part of the Dirac matrix $\gamma_{\mu}$,
\begin{equation}
  \Pi_{\pm}=\frac{{\cal M}\pm\hat{{\cal P}}}{2{\cal M}}\label{Pi}.
\end{equation}
Here, $\hat{{\cal P}}=\gamma^{\mu}{\cal P}_{\mu}$ and ${\cal M}=\sqrt{{\cal P}^2}$. The coefficients are
\begin{align}
  c_{1,2,3}=\frac{1}{\sqrt{\varepsilon_{q_{1,2,3}}+m}},\quad c_{N}=\frac{1}{\sqrt{\varepsilon_p+M}},\label{c123}\\
  \varepsilon_{q_{1,2,3}}=\sqrt{\vec{q}_i^2+m}={\cal P}\cd k_i/{\cal M},
\quad \varepsilon_{p}={\cal P}\cd p/{\cal M}.
\end{align}
As is well known, the bilinear form $\bar{u}\gamma_{\mu}u$ transforms as a four-vector under rotation and Lorentz transformation. 
According to (\ref{uPp}), the spinors in Eqs.~(\ref{sigmas1}) and (\ref{sigmas2}) are subjected upon by the rotation $R(g,{\cal P})$, and therefore, the vectors $\vec{\sigma}_{12}$ and $\vec{\sigma}_{3N}$ inherit the same transformation properties as the vectors $\vec{q}_i$ and $\vec{n}$.

Owing to the presence of $U_c$ in the definition of $\vec{\sigma}_{12}$ and of $\gamma_5$ in the definition of $\vec{\sigma}_{3n}$, these are actually pseudovectors. 
$\vec{\sigma}_{12}^{\sigma_1\sigma_2}$ with respect to the indices $\sigma_1,\sigma_2$, and $\vec{\sigma}_{3N}^{\sigma_3\sigma}$ with respect to $\sigma_3,\sigma$ are the $2\times 2$ matrices. 
It is easy to verify that at $\vec{\cal P}=0$ they indeed coincide with the Pauli matrices $\vec{\sigma}$ sandwiched with the corresponding spinors:
\begin{align*}
  \vec{\sigma}_{12}\vert_{\vec{\cal P}=0}=(w_1^{\dagger}\vec{\sigma}\sigma_y w_2^*),\quad 
  \vec{\sigma}_{3N}\vert_{\vec{\cal P}=0}=(w_3^{\dagger}\vec{\sigma}w_N).
\end{align*}
This correspondence justifies referring to $\vec{\sigma}_{12}$ and $\vec{\sigma}_{3N}$ as the ``relativistic generalization of the Pauli matrices'', and it enables to use them in the construction of the wave function similar to the ordinary Pauli matrices in Eqs.~(\ref{chinr_1}-\ref{chinr_5}). 

We will also need the following matrices, which serve as the counterparts of the unit operators:
\begin{align}
  &1_{12}=c_1c_2\bar{u}_{\cal P}^{\sigma_1}(k_1)\Pi_+ \gamma_5\Pi_- U_c \bar{u}_{\cal P}^{\sigma_2}(k_2)
  \vert_{\vec{\cal P}=0}=(w_1^{\dagger}\sigma_y w_2^*),\label{eq:11}
  \\
  &1_{3N}=c_3 c_N\bar{u}_{\cal P}^{\sigma_3}(k_3)\Pi_+ \Pi_+ U_c u_{\cal P}^{\sigma}(p)\vert_{\vec{\cal P}=0}=(w_3^{\dagger}w). \label{eq:12}
\end{align}

This construction has been realized in Ref.~\cite{karmNWF}, using, instead of the Jacobi momenta, the momenta $\vec{q}_i$ and $\vec{n}$ defined in Eqs.  (\ref{vqi}), (\ref{vn}). Below, we will repeat this construction in terms of the Jacobi momenta (\ref{Jacpqa}), which are more appropriate for comparison with the non-relativistic results of Ref.~\cite{baru}, where the wave functions are also presented in terms of the Jacobi momenta.

\subsection{Relativistic 3D basis}\label{3Da}

In terms of the variables $\vec{q}_i,\vec{n}$ (and the corresponding Jacobi momenta $\vec{k},\vec{q}$), as well as  the ``relativistic generalization of the Pauli matrices'' in Eqs.~(\ref{sigmas1}) and (\ref{sigmas2}) and the ``unit matrices'' in Eqs.~(\ref{eq:11}) and (\ref{eq:12}), the construction of the relativistic basis functions becomes similar to what we did in the non-relativistic case. 

The first five functions have exactly the same form as their non-relativistic counterparts given in Eqs.~(\ref{chinr_1}-\ref{chinr_5}), with the replacements 
the particle momenta $\vec{k}_i$, entering in the Jacobi momenta $\vec{q},\vec{k}$,
by the relativistic expressions $\vec{q}_i$, Eq.~(\ref{vqi}), and the matrices 
$(\sigma^i\sigma^y)_{\sigma_1\sigma_2}$ and $(\sigma^j)_{\sigma_3\sigma}$ by their relativistic counterparts $\vec{\sigma}_{12}^{\sigma_1\sigma_2}$ and 
$\vec{\sigma}_{3N}^{\sigma_3\sigma}$, Eqs.~(\ref{sigmas1}) and (\ref{sigmas2}).

We construct the following tensors:
\begin{align}
  T_{kk}^{ij}&=\left(\frac{k_{i}k_{j}}{\vec{k}^2}-\frac{1}{3}\delta_{ij}\right),\quad T_{qq}^{ij}=\left(\frac{q_{i}q_{j}}{\vec{q}^2}-\frac{1}{3}\delta_{ij}\right),
  \\
  T_{kq}^{ij}&=  \sum_{j^\prime }\left(\frac{k_{i}k_{j^\prime }}{\vec{k}^2}-\frac{1}{3}\delta_{ij^\prime }\right)\left(\frac{q_{j^\prime }q_{j}}{\vec{q}^2}-\frac{1}{3}\delta_{j^\prime j}\right), 
\end{align}
which appear in Eqs.~(\ref{chinr_1}-\ref{chinr_5}).
By means of these tensors, we rewrite the first five basis functions (\ref{chinr_1}-\ref{chinr_5}) in the form:
\begin{equation}\label{chi1-5}
  \begin{array}{ll}
    \chi_1=\frac{i}{4\pi\sqrt{2}}(\sigma^y)_{\sigma_1\sigma_2}\delta_{\sigma_3\sigma} &(A),
    \\
    \chi_2=-\frac{i}{4\pi\sqrt{6}}\vec{\sigma}_{12}\cd\vec{\sigma}_{3N} &(S),
    \\
    \chi_3=\frac{i\sqrt{3}}{8\pi}T_{kk}^{ij}\sigma^i_{12}\sigma^j_{3N}&(S),
    \\
    \chi_4=\frac{i\sqrt{3}}{8\pi}T_{qq}^{ij}\sigma^i_{12}\sigma^j_{3N}&(S),
    \\
    \chi_5=-\frac{3i\sqrt{6}}{16\pi}T_{kq}^{ij}\sigma^i_{12}\sigma^j_{3N}&(S).
  \end{array}
\end{equation}
The symbols ($A$) and ($S$) indicate the antisymmetric and symmetric properties of $\chi_{1-5}$ relative to the permutation $12\to 21$, respectively. The 
non-relativistic basis (\ref{chinr_1}-\ref{chinr_5}) contains just these five spin structures.

Another two tensors are constructed, using the first degree of the vector $\vec{n}$:
\begin{align}
  T_{kn}^{ij}&=\frac{1}{|\vec{k}|}\left(k_{i}n_{j}+n_{i}k_{j}-\frac{2}{3}\vec{k}\cd\vec{n}\delta_{ij}\right),
  \\
  T_{qn}^{ij}&=\frac{1}{|\vec{q}|}\left(q_{i}n_{j}+n_{i}q_{j}-\frac{2}{3}\vec{q}\cd\vec{n}\delta_{ij}\right).
\end{align}
We don't use the symmetric tensor 
\begin{align*}
  T_{nn}^{ij}=n_{i}n_{j}-\frac{1}{3}\delta_{ij},
\end{align*}
since there exists a relation among these tensors (Appendix B in Ref.~\cite{karmNWF}), which allows us to exclude one of them. Here, we choose to exclude $T_{nn}^{ij}$.

There are three antisymmetric tensors (relative to the permutation $i\leftrightarrow j$):
\begin{align}
  A^{ij}_{kn}&=\frac{1}{|\vec{k}|}(k_{i}n_{j}-k_{j}n_{i}),
  \quad
  A^{ij}_{qn}=\frac{1}{|\vec{q}|}(q_{i}n_{j}-q_{j}n_{i}),\\
  A^{ij}_{kq}&=\frac{1}{|\vec{k}||\vec{q}|}(k_{i}q_{j}-k_{j}q_{i}),
\end{align}

In further construction, we replace in $\chi_3,\chi_4$ the symmetric tensors $T_{kk}^{ij},T_{qq}^{ij}$ by  $T_{kn}^{ij},T_{qn}^{ij}$ and obtain new basis functions:
\begin{equation}\label{chi6-7}
  \begin{array}{ll}
  \chi_6=\frac{3i}{\sqrt{10}}\frac{1}{8\pi}T_{kn}^{ij}\sigma^i_{12}\sigma^j_{3N} & (A),
  \\
  \chi_7=\frac{3i}{\sqrt{10}}\frac{1}{8\pi}T_{qn}^{ij}\sigma^i_{12}\sigma^j_{3N} & (S).
  \end{array}
\end{equation}
Then we use antisymmetric tensors:
\begin{equation}\label{chi8-16}
  \begin{array}{ll}
    \chi_8=-\sqrt{\frac{3}{2}}\frac{1}{8\pi} A_{kq}^{ij}\sigma^i_{12}\sigma^j_{3N} &(A),
    \\
    \chi_9=-\sqrt{\frac{3}{2}}\frac{1}{8\pi}A_{kn}^{ij}\sigma^i_{12}\sigma^j_{3N} & (A),
    \\
    \chi_{10}=-\sqrt{\frac{3}{2}}\frac{1}{8\pi}A_{qn}^{ij}\sigma^i_{12}\sigma^j_{3N} & (S),
    \\
    \chi_{11}=\frac{\sqrt{3}}{8\pi}\vec{\sigma}_{12}\cd [\vec{k}\times\vec{q}]/(|\vec{k}||\vec{q}|) &(A),
    \\
    \chi_{12}=\frac{\sqrt{3}}{8\pi}\vec{\sigma}_{12}\cd [\vec{k}\times\vec{n}]/|\vec{k}|&(A),
    \\
    \chi_{13}=\frac{\sqrt{3}}{8\pi}\vec{\sigma}_{12}\cd [\vec{q}\times\vec{n}]/|\vec{q}| &(S),
    \\
    \chi_{14}=\frac{i}{8\pi}\vec{\sigma}_{3N}\cd [\vec{k}\times\vec{q}]/(|\vec{k}||\vec{q}|) &(S),
    \\
    \chi_{15}=\frac{i}{8\pi}\vec{\sigma}_{3N}\cd [\vec{k}\times\vec{n}]/|\vec{k}| &(S),
    \\
    \chi_{16}=\frac{i}{8\pi}\vec{\sigma}_{3N}\cd [\vec{q}\times\vec{n}]/|\vec{q}| &(A).
  \end{array}
\end{equation}
All these functions can be represented via the spherical functions and  Clebsch-Gordan coefficients, analogously to Eqs.~(\ref{chinr_1}-\ref{chinr_5}).

In this way, we have constructed 16 basis functions,  matching the number of the basis $V_{ij}$ constructed in Sec.~\ref{basisV}.
The full wave function, incorporating isospin, is again represented in the form (\ref{via_gij_f}).

In order to establish relation between the bases $V_{ij}$ in Eq.~(\ref{Vij}) and  $\chi_n$ in Eqs.~(\ref{chi1-5}-\ref{chi8-16}), we represent also the functions (\ref{chi1-5}-\ref{chi8-16})
in the ``standard'' representation, where the functions transform as in Eq.~(\ref{wfp6}) and   the basis (\ref{Vij}) was constructed. 
The explicit covariance will help us to solve this problem. 
If $\vec{\cal P}=0$, all the rotation operators $R$ in Eq.~(\ref{wfp7}) turn into unit operators, and hence, all the corresponding matrices $D^{(\frac{1}{2})}$ reduce to the unit matrices. Therefore, by evaluating the expressions at $\mathcal{P}=0$ in terms of the ``standard'' spinors and the Dirac matrices, we guess these expressions reduce to Eqs.~(\ref{chi1-5}-\ref{chi8-16}). 
These explicit expressions are given in Appendix \ref{appendix1}.

\subsection{Relation between the components $g_{ij}$ in (\ref{via_gij}) and $\psi_n$ in (\ref{via_psin})}

Both decompositions (\ref{via_gij}) and (\ref{via_psin}) represent {\it one and the same} wave function $\Phi_{12}$. Therefore,
$$
\sum_{i^\prime ,j^\prime =1}^{4}g_{i^\prime j^\prime }V_{i^\prime j^\prime }=\sum_{n=1}^{16}\psi_n\chi_n.
$$
Multiplying this equality on the left by $\frac{1}{2}V^{\dagger}_{ij}$ and using the orthogonality condition (\ref{Vijnorm}), 
we get a linear system of 16 equations for $\psi_n$:
\begin{equation}
  \sum_{n=1}^{16}M_{ij,n}\psi_n=g_{ij},\quad i,j=1,\cdots,4.
  \label{relat}
\end{equation}
where $M_{ij,n}=\frac{1}{2}V^{\dagger}_{ij}\chi_n$.
We rewrite $\chi_n$ in Eqs.~(\ref{chi1-5}-\ref{chi8-16}) as 
\begin{equation}
  \chi_{n}= c_{n}\sum_{\kappa} [\bar{u}(k_1)O^{(A\kappa)}_{n}U_c\bar{u}(k_2)]\,[\bar{u}(k_3)O^{(B\kappa)}_{n}u(p)].\label{chi_n}
\end{equation}
with the matrices $O^{(A\kappa)}_n$, $O^{(B\kappa)}_n$, and the coefficients $c_n$ given in Appendix~\ref{appendix1}.
According to the definitions of $\chi_{1-16}$, $\kappa=1$ for $n=1,2$ (no sum over $\kappa$ since it contains one term only), $\kappa=1,2$ for $n=3,4$, $\kappa=1,2,3,4$ for $n=5$, $\kappa=1,2,3$ for $n=6,7$, $\kappa=1,2$ for $n=8,9,10$, and again $\kappa=1$ for $n=11,\ldots,16$.
In this way, for the elements $M_{ij,n}$ we find:
\begin{align}
  M_{ij,n}&=\frac{1}{2}V^{\dagger}_{ij}\chi_n\nonumber\\
  &=\frac{1}{2}c_nC_{ij}\sum_{\kappa}\Tr[O_n^{(A\kappa)}(\hat{k}_2-m)\bar{T}_i(\hat{k}_1+m)]\nonumber\\
  &\times \Tr[O_n^{(B\kappa)}(\hat{p}+M)\bar{S}_j(\hat{k}_3+m)].\label{M}
\end{align}
$\bar{S}_j$ is defined in Eqs.~(\ref{eq:S3bar_1}-\ref{eq:S3bar_3}), $\bar{T}_i$ is defined in Eqs.~(\ref{Tbar1}) and (\ref{Tbar2}), $C_{ij}$ is defined in  (\ref{wf6}), (\ref{Cij}).

\section{Three-body bound state equation}\label{3beq}
\begin{figure}
  \begin{center}
  \mbox{\epsfxsize=9cm\epsffile{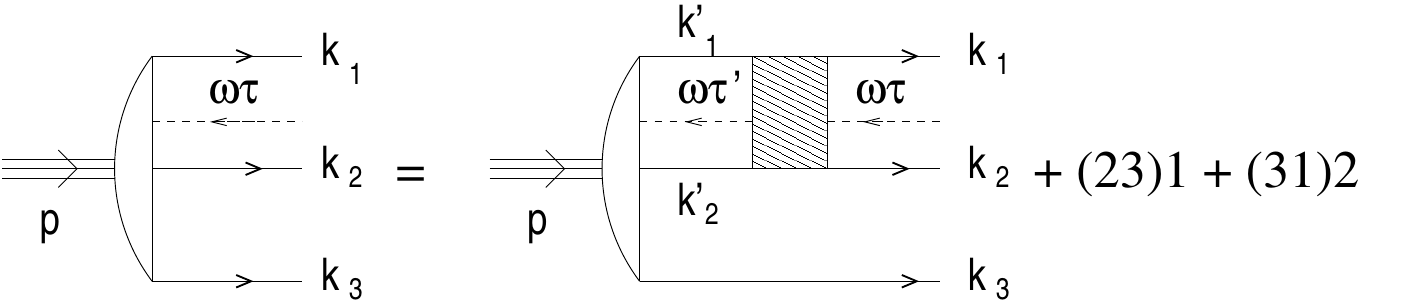}}
  \caption{Three-body equation for the vertex function $\Gamma$.\label{feq3b}}
  \end{center}
\end{figure}

In the present paper, the relativistic description of $^3$He is based on the three-body LF equation. 
Within the framework of explicitly covariant LFD graph techniques,  this equation for the vertex function $\Gamma$ is an element of the graph techniques, which is related to the wave function by Eq.~(\ref{Phi-Gam}), shown graphically for the two-body pair interaction in Fig.~\ref{feq3b}.
It is convenient to formally attribute the difference $k_1+k_2+k_3-p=\omega\tau$, which arises due to the non-conservation of the four-momenta (specifically, only the minus component in the standard LFD), to a fictive particle, called a spurion.
The dashed lines in Fig.~\ref{feq3b} correspond to the ``spurion''  carrying the momenta $\omega\tau$ or $\omega\tau’$.
The spurion momentum already appears in Eq.~(\ref{Psi}) as an argument of the wave function. 
The terms (23)1 and (31)2 correspond to  the contributions from the interacting pairs (23) and (31). 
Applying  the rules of the covariant LF graph techniques~\cite{cdkm} to Fig.~\ref{feq3b}, we find:
\begin{widetext}
  \begin{align}
    \Gamma(k_1,k_2,k_3,p,\omega\tau)&=
    \int \Gamma(k^\prime _1,k^\prime _2,k_3,p,\omega\tau^\prime )
    {\cal K}(k^\prime _1,k^\prime _2,\omega\tau^\prime ;k_1,k_2,\omega\tau)
    \delta^{(4)}(k^\prime _1+k^\prime _2-\omega\tau^\prime -k_1-k_2+\omega\tau)\frac{\mathrm{d} \tau^\prime }{\tau^\prime }
    \nonumber\\
    &\times
    \frac{1}{(2\pi)^3}\delta({k^\prime }^2_1-m^2)\mathrm{d}^4k^\prime _1
    \delta({k^\prime }^2_2-m^2)\mathrm{d}^4k^\prime _2 
    +(31)2+(23)1,\label{eq3}
  \end{align}
\end{widetext}
where ${\cal K}$ is the kernel of interaction between the particles 12. For the present, we omit the spin indexes. 

The relation between the vertex and wave functions reads:
\begin{equation}
  \Phi=\frac{\Gamma}{{\cal M}^2-M^2},\label{Phi-Gam}
\end{equation}
where 
\begin{equation}
  {\cal M}^2=(k_1+k_2+k_3)^2=\sum_{i=1}^3\frac{\vec{R}_{i\perp}^2+m^2}{x_i}.  \label{calM}
\end{equation}
Here, $M$ is the bound state mass, the definition of the LF variables $\vec{R}_{i\perp}$ and $x_i$, as well as the proof of the equality (\ref{calM}), are given in Appendix \ref{app1}.  Evaluating the integral on r.h.-side of Eq.~(\ref{eq3}), we obtain the following equation, 
\footnotesize
\begin{widetext}
  \begin{align}
    &({\cal M}^2-M^2)\Phi(\vec{R}_{1\perp},x_1,\vec{R}_{2\perp},x_2,\vec{R}_{3\perp},x_3)
    =
    \int \Phi(\vec{R^\prime }_{1\perp},x^\prime _1,\vec{R^\prime }_{2\perp},x^\prime _2,
    \vec{R^\prime }_{3\perp},x^\prime _3)
    {\cal K}(\vec{R^\prime }_{1\perp},x^\prime _1,\vec{R^\prime }_{2\perp},x^\prime _2;
    \vec{R}_{1\perp},x_1,\vec{R}_{2\perp},x_2)
    \frac{1}{(2\pi)^3}
    \nonumber\\
    &\times
    \delta^{(2)}\left(\sum_{i=1}^3 \vec{R^\prime }_{i\perp}\right)
    \delta\left(\sum_{i=1}^3 x^\prime _i-1\right)
    \delta^{(2)}\left(\vec{R^\prime }_{3\perp}- \vec{R}_{3\perp}\right)
    \;2x^\prime _3\delta(x^\prime _3-x_3)
    2\prod_{i=1}^3\frac{\mathrm{d}^2R^\prime _{i\perp}\mathrm{d}x^\prime _i}{2x^\prime _i}
    +(31)2+(23)1. \label{eq6}
  \end{align}
\end{widetext}
\normalsize
In the graph shown in Fig.~\ref{feq3b}, the particle No.~3 does not interact, and its momentum remains unchanged. This corresponds to the appearance of the delta functions $\delta^{(2)}(\vec{R^\prime }_{3\perp}- \vec{R}_{3\perp})2x^\prime _3\delta(x^\prime _3-x_3)$ in Eq.~(\ref{eq6}).

\subsection{Equations for the Faddeev components}\label{eqs_Fadd}

To find the solution of Eq.~(\ref{eq6}), we will use the standard method applied to any three-body system of identical particles with the pair interaction (see \eg,~ \cite{green}).
The wave function is represented by Eq.~(\ref{Psi1}) as a sum of three Faddeev components. Then, Eq.~(\ref{eq6}) turns into a system of equations for these components. However, each equation contains the interaction between only one pair of particles: $K_{12}$, $K_{23}$, or $K_{31}$. 
For identical particles, the Faddeev components are related to each other by permutations, so we get an equation for one Faddeev component (though with non-permuted and permuted arguments in one equation), with one pair interaction only. Symbolically, this reads:
\begin{align}
  ({\cal M}^2&-M^2)\Phi_{12}(1,2,3)=-K_{12}\cdot\Phi_{12}(1,2,3)
  \nonumber\\
  &-K_{12}\cdot\Phi_{12}(2,3,1)-K_{12}\cdot\Phi_{12}(3,1,2). \label{eq_symb}
\end{align}
Note that only the component $\Phi_{12}$ is subjected to permutation, whereas the kernel $K_{12}$ remains the same in all three terms on the r.h.-side,
or more specifically,
\begin{widetext}
  \begin{align}\label{eq6bcd}
    &({\cal M}^2-M^2)\Phi^{\sigma}_{12;\sigma_1\sigma_2\sigma_3}(\vec{R}_{1\perp},x_1,\vec{R}_{2\perp},x_2,\vec{R}_{3\perp},x_3)
    \nonumber\\
    =&-\sum_{\sigma^\prime _1,\sigma^\prime _2}\int 
    \frac{\mathrm{d}^2R^\prime _{2\perp}\mathrm{d}x^\prime _2}{(2\pi)^32x^\prime _1x^\prime _2} 
    \Phi^{\sigma}_{12;\sigma^\prime _1\sigma^\prime _2\sigma^{\phantom{^\prime }}_3}(\vec{R^\prime }_{1\perp},x^\prime _1,\vec{R^\prime }_{2\perp},x^\prime _2,\vec{R}_{3\perp},x_3)
    {\cal K}_{12,\sigma_1\sigma_2}^{\sigma^\prime _1\sigma^\prime _2}(\vec{R}_{1\perp},x_1,\vec{R}_{2\perp},x_2; \vec{R^\prime }_{1\perp},x^\prime _1,\vec{R^\prime }_{2\perp},x^\prime _2)
    \nonumber\\
    &
    -\sum_{\sigma^\prime _1,\sigma^\prime _2}\int 
    \frac{\mathrm{d}^2R^\prime _{2\perp}\mathrm{d}x^\prime _2}{(2\pi)^32x^\prime _1x^\prime _2}
    \Phi^{\sigma}_{12;\sigma^\prime _2\sigma_3\sigma^\prime _1}(\vec{R^\prime }_{2\perp},x^\prime _1,\vec{R}_{3\perp},x_3,\vec{R^\prime }_{1\perp},x^\prime _1)
    {\cal K}_{12,\sigma_1\sigma_2}^{\sigma^\prime _1\sigma^\prime _2}(\vec{R}_{1\perp},x_1,\vec{R}_{2\perp},x_2; \vec{R^\prime }_{1\perp},x^\prime _1,
    \vec{R^\prime }_{2\perp},x^\prime _2)
    \nonumber\\
    &
    -
    \sum_{\sigma^\prime _1,\sigma^\prime _2}\int 
    \frac{\mathrm{d}^2R^\prime _{2\perp}\mathrm{d}x^\prime _2}{(2\pi)^3 2x^\prime _1x^\prime _2}
    \Phi^{\sigma}_{12;\sigma_3\sigma^\prime _1\sigma^\prime _2}(\vec{R}_{3\perp},x_3,\vec{R^\prime }_{1\perp},x^\prime _1,\vec{R^\prime }_{2\perp},x^\prime _2)
    {\cal K}_{12,\sigma_1\sigma_2}^{\sigma^\prime _1\sigma^\prime _2}(\vec{R}_{1\perp},x_1,\vec{R}_{2\perp},x_2; \vec{R^\prime }_{1\perp},x^\prime _1,\vec{R^\prime }_{2\perp},x^\prime _2)
  .
  \end{align}
\end{widetext}
Equation  (\ref{eq6bcd}) implies also summation over the isospin indexes
omitted for shortness.

\subsection{$NN$ kernel}\label{Kernel}

\begin{table}
  \begin{center}
    \caption{\scriptsize Parameters of the exchanged mesons~\cite{M1}. The data for $\sigma_0$ and $\sigma_1$ are given for exchanges in the $NN$ states with the isospins $T=0$ and $T=1$ correspondingly.}\label{tab3}
    \begin{tabular}{ccccccc}
      \hline\hline
      \noalign{\smallskip} 
        &$J^{\pi}$  &   $T$  & $\mu$ (MeV) & $g^2/(4\pi)\; [f/g]$  & $\Lambda$ (GeV) & $n$  \\
      \noalign{\smallskip}\hline 
      \noalign{\smallskip}
      $\pi$ & $0^-$ &1& 138.03   &   14.6      &1.3&1  \\ 
      $\eta$& $0^-$ &0 & 548.8 &   5.0  & 1.5    &1    \\
      $\delta$ & $0^+$ &1 & 983 &    1.1075    & 2 &1    \\
      $\sigma_0$  & $0^+$& &720 &16.9822 & 2 & 1 \\
      $\sigma_1$ & $0^+$ & &  550&8.2797    & 2  & 1   \\
      $\omega$& $1^-$ &0  & 782.6&  20.0\; [0.0]        & 1.5 &1  \\
      $\rho$ & $1^-$ & 1&769&  0.81\; [6.1]  &   2    &  2   \\
      \noalign{\smallskip}\hline\hline
    \end{tabular}\\
  \end{center}
\end{table}

For the interaction kernel, we take the one-boson exchange model determined by  the exchanges by the pseudoscalar ($\pi$, $\eta$), scalar ($\delta$, $\sigma_0$, $\sigma_1$), and vector ($\omega$, $\rho$) mesons. 
These mesons and their parameters, taken from \cite{M1}, are listed in  Table~\ref{tab3}.
Here, $\sigma_0$ and $\sigma_1$ are the phenomenological (fictive)  mesons. The  $\sigma_0$ meson exchange takes place in the NN state with the isospin 0, while the $\sigma_1$ meson exchange takes place in the NN state with the isospin 1. 

\begin{figure}
  \begin{center}
    \mbox{\epsfxsize=9cm\epsffile{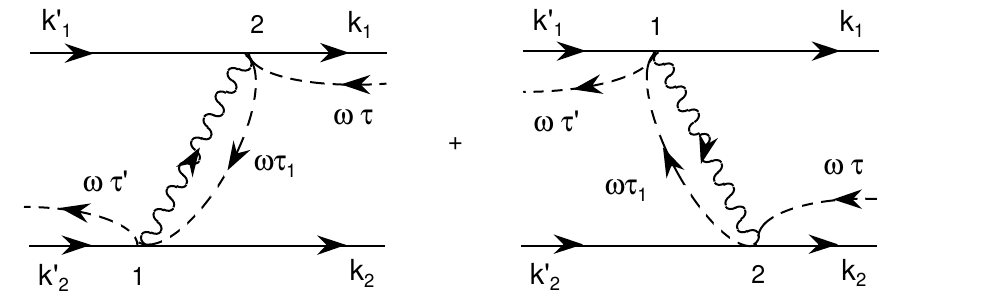}}
    \caption{One boson exchange kernel.\label{fkern}}
  \end{center}
\end{figure}

In the framework of explicitly covariant LFD, the kernel is graphically shown in Fig.~\ref{fkern}. It has two contributions
corresponding to the two time-ordered diagrams (in the LF time).
Applying the rules of the explicitly covariant LFD~\cite{cdkm} to the graph Fig. \ref{fkern}, we get
for the S and PS couplings:
\begin{equation}
{\cal K}^{\sigma^\prime _2\sigma^\prime _1}_{\sigma_2\sigma_1}
  =\Pi_{12}(Q^2) \Bigl[\bar{u}^{\sigma_1}(k_1)O_1 u^{\sigma^\prime _1}(k^\prime _1)\Bigr]\,
  \Bigl[\bar{u}^{\sigma_2}(k_2)O_2 u^{\sigma^\prime _2}(k^\prime _2)\Bigr],\label{calK}
\end{equation}
where $\Pi_{12}(Q^2)$ is the scalar part of the propagator. 
Its explicit form is given in Eq.~(\ref{Pi12}). 
For the scalar exchange: $O_1=O_2=g_{\rm s}$. 
For the pseudoscalar exchange: $O_1=O_2=i\gamma_5 g_{\rm{ps}}$.  
For the exchange by the meson with the isospin $T=1$, the product of these matrices also contains the factor $\vec{\tau}_1\cd\vec{\tau}_2$. 
If the isospin of the exchanged meson is 0, then the factor $\vec{\tau}_1\cd\vec{\tau}_2$ is replaced by 1.

The kernel for the vector coupling reads
\begin{align}
  K^{\sigma^\prime _2\sigma^\prime _1}_{\sigma_2\sigma_1}
  &=\Pi_{12}(Q^2) L_{\alpha\beta}\Bigl[\bar{u}^{\sigma_1}(k_1)O_1^{\alpha}
  u^{\sigma^\prime _1}(k^\prime _1)\Bigr]
  \nonumber\\
  &\times\Bigl[\bar{u}^{\sigma_2}(k_2)O_2^{\beta} u^{\sigma^\prime _2}(k^\prime _2)\Bigr],\label{v3}
\end{align}
with
\footnotesize
\begin{align}
  L_{\alpha\beta}=&-g_{\alpha\beta} + \frac{\theta[\omega\cd(k_1-k^\prime _1)]}{\mu^2}(k_1-k^\prime _1-\omega\tau)_{\alpha}(k^\prime _2-k_2-\omega\tau^\prime )_{\beta}
  \nonumber\\
  &\phantom{a}
  \nonumber\\
  &+\frac{\theta[\omega\cd(k^\prime _1-k_1)]}{\mu^2}(k^\prime _1-k_1-\omega\tau^\prime )_{\alpha}
  (k_2-k^\prime _2-\omega\tau)_{\beta}, 
  \label{v4}
\end{align}
and vertex operators
\begin{align}
  &O_1^{\alpha}=\left[g_{\rm v}\gamma^{\alpha}+
  \frac{\displaystyle{f}}{\displaystyle{2m}}\sigma^{\alpha^\prime \alpha}(-i)
  (k_1-k^\prime _1-\omega\tau)_{\alpha^\prime }\right]\theta[\omega\cd(k_1-k^\prime _1)]
  \nonumber\\
  &+\left[g_{\rm v}\gamma^{\alpha}+\frac{\displaystyle{f}}{\displaystyle{2m}}\sigma^{\alpha^\prime \alpha}(i)
  (k^\prime _1-k_1-\omega\tau^\prime )_{\alpha^\prime }\right]\theta[\omega\cd(k^\prime _1-k_1)],
  \label{v5}\\
  &O_2^{\beta}=\left[
  g_{\rm v}\gamma^{\beta}+\frac{\displaystyle{f}}{\displaystyle{2m}}\sigma^{\beta^\prime \beta}(i)
  (k^\prime _2-k_2-\omega\tau^\prime )_{\beta^\prime }\right]
  \theta[\omega\cd(k_1-k^\prime _1)]
  \nonumber\\
  &+\left[g_{\rm v}\gamma^{\beta}+\frac{\displaystyle{f}}{\displaystyle{2m}}\sigma^{\beta^\prime \beta}(-i)
  (k_2-k^\prime _2-\omega\tau)_{\beta^\prime }\right]
  \theta[\omega\cd(k^\prime _1-k_1)].
  \label{v6}
\end{align}
\normalsize
For regularization, we multiply each vertex by the form factor:
\begin{equation}
  F(Q^2) =\left(\frac{\Lambda^2-\mu^2}{\Lambda^2+ Q^2}\right)^n,\label{BFF}
\end{equation}
where $\Lambda$ and $n$ are parameters whose values depend on the coupling and are given in Table~\ref{tab3}. The argument $Q^2$ is the same as in 
$\Pi_{12}(Q^2)$, and it is defined in Eq.~(\ref{Q2}).

We substitute the wave function in Eq.~(\ref{eq6bcd}) using the decomposition in Eqs.~(\ref{via_gij}) and (\ref{via_gij_f}) in the spin-isospin basis $V_{ij}\xi^{(0,1)}$, where $V_{ij}$ and $\xi^{(0,1)}$ defined in (\ref{Vij}) and (\ref{eq:xi01a_1}-\ref{eq:xi01a_2})
respectively. Then, we multiply both sides of the equation (\ref{eq6bcd}) by $V_{ij}^{\dagger}\xi^{(0,1)\tau}_{\tau_1\tau_2\tau_3}$ and take sum over the spin-isospin projections.
As a result, this equation, being rewritten for the components $g^{(0,1)}_{12;ij}\equiv g^{(t)}_{12;ij}$ ($t=0,1$), obtains the form: 
\footnotesize
\begin{align}
  &({\cal M}^2-M^2)g^{(t)}_{12;ij}(1,2,3) \label{gij}
  \\
  &=
  \sum_{i^\prime j^\prime t^\prime }  \int \frac{\mathrm{d}^2R^\prime _{2\perp}\mathrm{d}x^\prime _2}{(2\pi)^32x^\prime _1x^\prime _2}   g^{(t^\prime)}_{12;i^\prime j^\prime }(1^\prime ,2^\prime ,3) W_{ijt}^{i^\prime j^\prime t^\prime }\mbox{\tiny (12)}(1^\prime ,1;2^\prime ,2;3) 
  \nonumber\\
  &+
  \sum_{i^\prime j^\prime t^\prime }  \int \frac{\mathrm{d}^2R^\prime _{2\perp}\mathrm{d}x^\prime _2}{(2\pi)^32x^\prime _1x^\prime _2}  g^{(t^\prime)}_{12;i^\prime j^\prime }(3,1^\prime ,2^\prime )W_{ijt}^{i^\prime j^\prime t^\prime }\mbox{\tiny (31)}(1^\prime ,1;2^\prime ,2;3) 
  \nonumber\\
  &+
  \sum_{i^\prime j^\prime t^\prime } \int \frac{\mathrm{d}^2R^\prime _{2\perp}\mathrm{d}x^\prime _2}{(2\pi)^32x^\prime _1x^\prime _2} g_{12;i^\prime j^\prime }^{(t^\prime)}(2^\prime ,3,1^\prime )W_{ijt}^{i^\prime j^\prime t^\prime }\mbox{\tiny (23)}(1^\prime ,1;2^\prime ,2;3). \nonumber
\end{align}
\normalsize
Each index $i,j,i^\prime,j^\prime $ runs over the values $1,2,3,4$, and each pair isospin index $t,t^\prime$ takes the values $0,1$. 

Each permuted kernel in Eq.~(\ref{gij}) is given by the sum of exchanges involving the seven mesons listed in Table~\ref{tab3}:
\begin{align*}
  W_{ijt}^{i^\prime j^\prime t^\prime }=\sum_{n=\rm{mesons}}W_{ijt}^{i^\prime j^\prime t^\prime }(n).
\end{align*}
The permuted kernels read
\begin{widetext}
 \begin{align}
    W_{ijt}^{i^\prime j^\prime t^\prime }\mbox{\tiny (12)}&=-\Pi_{12}F^2(Q^2)\eta(123,123;t,t^\prime ;I)
    C_{ij}(k_1,k_2)C_{i^\prime j^\prime }(k^\prime _1,k^\prime _2)
    \nonumber\\
    &\times \Tr[\bar{T}_{i}(1,2,3)(\hat{k}_1+m)O_1(\hat{k}^\prime _1+m)T_{i^\prime }(1^\prime ,2^\prime ,3)(-\hat{k}^\prime _2+m)\tilde{O}_2(-\hat{k}_2+m)]\delta_{jj^\prime },
    \label{W12}\\
    W_{ijt}^{i^\prime j^\prime t^\prime }\mbox{\tiny (31)}&=\frac{1}{2}\Pi_{12}F^2(Q^2)\eta(123,312;t,t^\prime ;I)C_{ij}(k_1,k_2)C_{i^\prime j^\prime }(k_3,k^\prime _1)
    \nonumber\\
    &\times \Tr\Bigl[(\hat{k}_2+m)O_2(\hat{k^\prime }_2+m)S_{j^\prime }(2^\prime )(\hat{p}+M)\bar{S}_j(3)(\hat{k}_3+m)T^\prime _{i^\prime }(3,1^\prime ,2^\prime )
    \nonumber\\
    &\times(-\hat{k^\prime }_1+m)\tilde{O}_1 (-\hat{k}_1+m) \tilde{\bar{T}}_i (1,2,3) \Bigr],
    \label{W31}\\
    W_{ijt}^{i^\prime j^\prime t^\prime }\mbox{\tiny (23)}&=\frac{1}{2}\Pi_{12}F^2(Q^2)\eta(123,231;t,t^\prime ;I)C_{ij}(k_1,k_2)C_{i^\prime j^\prime }(k^\prime _2,k_3)
    \nonumber\\
    &\times \Tr\left[\bar{T}_{i}(1,2,3)(\hat{k}_1+m)O_1 (\hat{k^\prime }_1+m)S^\prime _{j^\prime }(1^\prime ) (\hat{p}+M)
    \bar{S}_{j}(3) (\hat{k}_3+m)\right.
    \nonumber\\
    &\times\left.\tilde{T}_{i^\prime }(2^\prime ,3,1^\prime )(-\hat{k}^\prime _2+m)\tilde{O}_2(-\hat{k}_2+m)\right],
    \label{W23}
  \end{align}
\end{widetext}
where the factors $\eta$ are given in the Table \ref{tab2}. $\tilde{O}=U_c O^t U_c$ ($O^t$ is the transposed matrix), so that
\begin{align*}
  & \tilde{1}=1,\quad
  \tilde{\gamma}^{\mu}=-\gamma^{\mu},\quad \widetilde{\gamma^{\mu}\gamma^{\nu}}=\gamma^{\nu}\gamma^{\mu},
  \\
  & \tilde{\gamma}_5=\gamma_5,\quad
  \widetilde{\gamma^{\mu}\gamma_5}=\gamma^{\mu}\gamma_5.
\end{align*}

According to Eq.~(\ref{eq_symb}), 
the factors $\Pi_{12}$ and $F(Q^2)$ are the same in  all three equations (\ref{W12}-\ref{W23}). In the system of equations for the Faddeev components, only the components are subjected to permutations. $F(Q^2)$ is defined in Eq.~(\ref{BFF}), whereas $\Pi_{12}$ reads
\begin{equation}
  \Pi_{12}=\frac{1}{\mu^2+Q^2},\label{Pi12}
\end{equation}
where
\begin{widetext}
  \begin{eqnarray}
    Q^2&=&
    \left\{
    \begin{array}{ll}
    -(k_1-k^\prime _1)^2+2\tau \omega\cd(k_1-k^\prime _1), & \mbox{if  $\omega\cd(k_1-k^\prime _1)\geq 0$}
    \\
    -(k^\prime _1-k_1)^2 +2\tau^\prime  \omega\cd (k^\prime _1-k_1), & \mbox{if  $\omega\cd(k_1-k^\prime _1)\leq 0$}
    \end{array}
    \right.
    \nonumber\\
    &=&
    \left\{
    \begin{array}{ll}
    \Bigl({\cal M}^2-M^2\Bigr)(x_1-x^\prime _1)+\frac{m^2(x_1-x^\prime _1)^2+(x_1\vec{R^\prime }_{1\perp}-x^\prime _1\vec{R}_{1\perp})^2}{x_1x^\prime _1},&
    \mbox{if  $x_1-x^\prime _1\geq 0$}
    \\
    \Bigl({{\cal M}^\prime }^2-M^2\Bigr)(x^\prime _1-x_1)+\frac{m^2(x_1-x^\prime _1)^2+(x_1\vec{R^\prime }_{1\perp}-x^\prime _1\vec{R}_{1\perp})^2}{x_1x^\prime _1},&
    \mbox{if  $x_1-x^\prime _1\leq 0$}
    \end{array}
    \right., 
    \label{Q2}
  \end{eqnarray}
\end{widetext}
The four-vector scalar products in Eq.~(\ref{Q2}) are expressed in terms of the variables $\vec{R}_i$ and $x_i$ using the relations established in Appendix~\ref{kinemat}.
${\cal M}^2$ is defined in Eq.~(\ref{calM}), and 
\begin{align*}
  {\cal M^\prime }^2=\frac{\vec{R^\prime }_{1\perp}^2+m^2}{x_1^\prime}+ \frac{\vec{R^\prime }_{2\perp}^2+m^2}{x^\prime _2}+ \frac{\vec{R}_{3\perp}^2+m^2}{x_3}.
\end{align*}

\section{Numerical results}\label{num}

Because the wave function in the integral equation~(\ref{eq6bcd}) depends on five variables, a direct numerical solution is rather challenging. 
To address this, we employ an iterative approximation scheme, which was successfully applied in Ref.~\cite{ck95} to determine the deuteron's LF wave function, to obtain the relativistic $^3$He wave functions.
Beginning with the non-relativistic solution $\psi_{\mathrm{nr},n}^{(0,1)}$ from Ref.~\cite{baru}, we extract the components $g_{\mathrm{nr},ij}^{(0,1)}$ via Eq.~(\ref{relat}), insert them into the r.h.-side of Eq.~(\ref{eq6bcd}), and evaluate the resulting integral. 
After imposing the normalization condition in Eqs.~(\ref{norm2}-\ref{norm2_3}), this first iteration yields the approximate relativistic scalar functions $g_{\mathrm{r},ij}^{(0,1)}$. 
Finally, we construct the corresponding relativistic wave functions $\psi_{\mathrm{r},n}^{(0,1)}$ determining via Eq.~(\ref{relat}).

\begin{table}
  \begin{center}
    \caption{
      Comparison of the proportion of the pair isospin components for non-relativistic and relativistic wave functions. The normalization formulas are given in Eqs.~(\ref{norm2_2}) and (\ref{norm2_3}).
    }
    \begin{tabular}{c|ccc}
      \hline\hline
        & $I^{(0)}$ & $I^{(1)}$ & $I_{\mathrm{mix}}$ \\ 
    \hline
    non-rel. & 0.365  & 0.090  & 0.545  \\
    rel. & 0.180  &  0.294 & 0.526  \\ 
    \hline\hline
    \end{tabular}
    \label{norm_res}
  \end{center}
\end{table}

By Eqs.~(\ref{norm2_2}) and (\ref{norm2_3}), we calculate the proportion of the pair isospin components. 
The three-body normalization integral in Eq.~(\ref{norm1}) contains three permuted terms.
Our calculation reveals that the integrals for the terms obtained by the cyclic permutations are identical within a certain numerical error ($< 0.1\%$), which is consistent with the properties of identical particles.
Table~\ref{norm_res} shows the comparison of the proportion for non-relativistic and relativistic wave functions. 
Our results show that the mixed isospin permutation term $I_{\mathrm{mix}}$ is dominant regardless of whether relativistic effects are considered or not.
Besides, the proportion of the relativistic component with the pair isospin $t=1$ is significantly larger than that of the non-relativistic component, implying the influence of relativistic corrections on isospin components.

\subsection{$q$-dependence}\label{q}
As mentioned, the relativistic $^3$He LF wave function is determined by a set of 32 spin-isospin components, either $g^{(t)}_{ij}$ ($i,j=1,\ldots, 4$, $t=0,1$),
or $\psi^{(t)}_n$ ($n=1,\ldots, 16$, $t=0,1$), which are related to each other by Eq.~(\ref{relat}).  
Each of the components depends on five variables, which can be chosen as: $\psi=\psi(\vec{q}_1,\vec{q}_2,\vec{q}_3,\vec{n})$, where $\vec{q}_i,\vec{n}$ are defined in Eqs.~(\ref{vqi}) and (\ref{vn}), respectively.
By definition, $\vec{q}_1+\vec{q}_2+\vec{q}_3=0$. 
The function $\psi(\vec{q}_1,\vec{q}_2,\vec{q}_3,\vec{n})$, being a scalar function, depends, say, on five scalars 
$\vec{q}_1^2$, $\vec{q}_2^2$, $\vec{q}_1\cd\vec{q}_2$, $\vec{q}_1\cd\vec{n}$, and $\vec{q}_2\cd\vec{n}$.

Since, due to the conservation law (\ref{sumqi}), the momenta $\vec{q}_i$ lie in the same plane, we choose a configuration, when $q_1=q_2=q_3=q$, and the angle between these vectors is 120 degrees. Additionally, we take $\vec{n}$ to be orthogonal to the plane formed by the momenta $\vec{q}_i$. 
In this configuration, all the components depend on $q$ only. 

\begin{figure}
  \centering
  \includegraphics[width=0.94\linewidth]{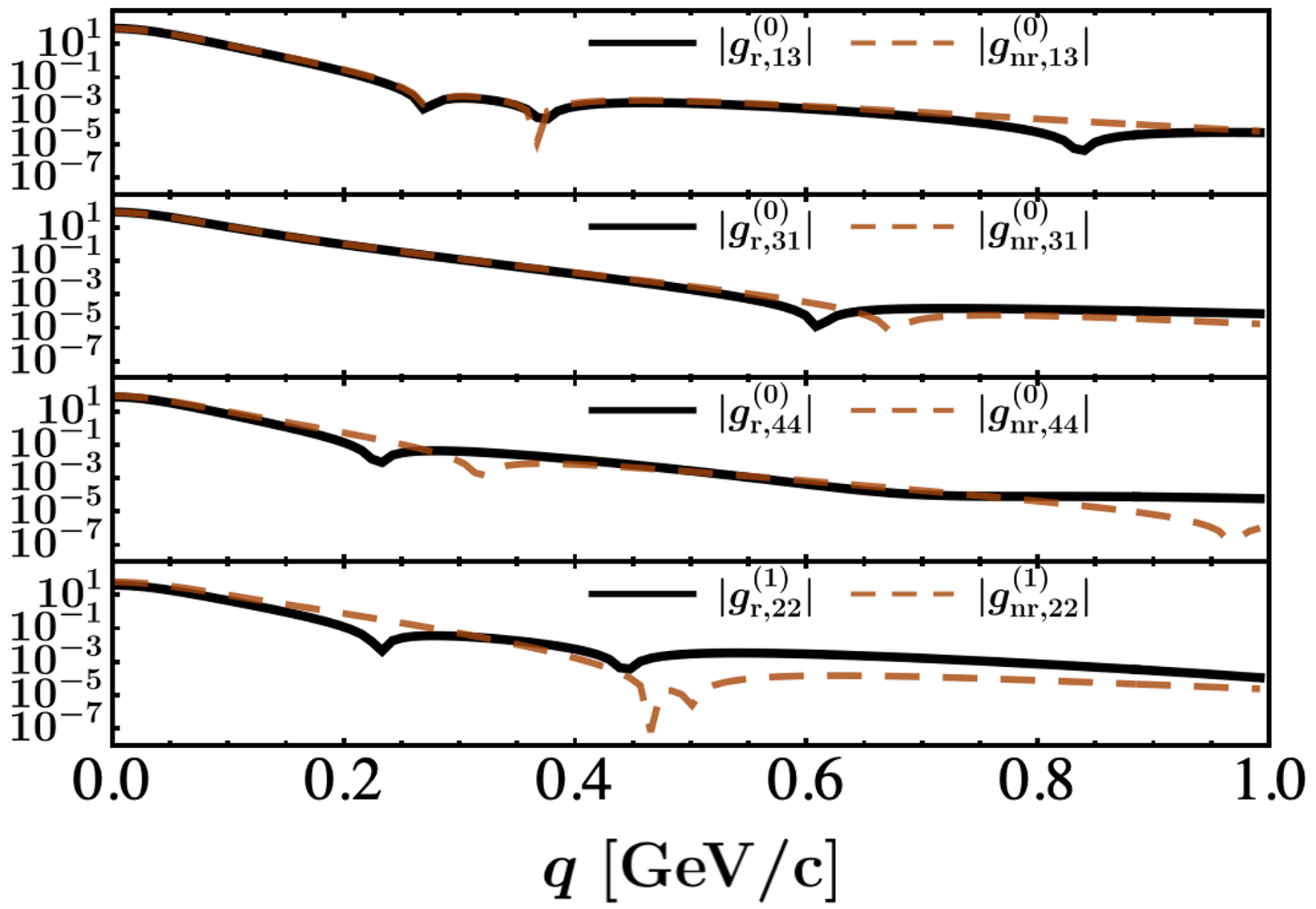}
  \caption{
    Logarithmic plots of dominant components (dashed lines: non-relativistic components, solid lines: relativistic components) as functions of $q$. 
    The fixed arguments are $x_1=x_2=x_3=1/3$, $q_1 = q_2 = q_3 = q$, and $\langle \vec{q}_1, \vec{q}_2\rangle = \langle \vec{q}_1, \vec{q}_3\rangle = \langle \vec{q}_2, \vec{q}_3\rangle = 2\pi/3$.
    The component functions are in units of $\mathrm{GeV^{-3}}$.
  }
  \label{Fig:qqq120}
\end{figure}

As explained, the $^3$He LF wave function comprises 32 scalar function components. In the numerical calculations, we put the nucleon mass $m=1$ GeV.
For clarity, Fig.~\ref{Fig:qqq120} presents four dominant components, three components for $t=0$ and one component for $t=1$ dominate, as functions of $q$.
Each of these dominant components at $q=0$ has a magnitude of the order $10$ GeV$^{-3}$. 
In the non-relativistic domain (at small $q\ll m$), the relativistic components align closely with their non-relativistic counterparts.
The deviations of the dominant components only appear once $q$ grows into the relativistic domain due to relativistic effects.

\begin{figure}
  \centering
  \includegraphics[width=0.98\linewidth]{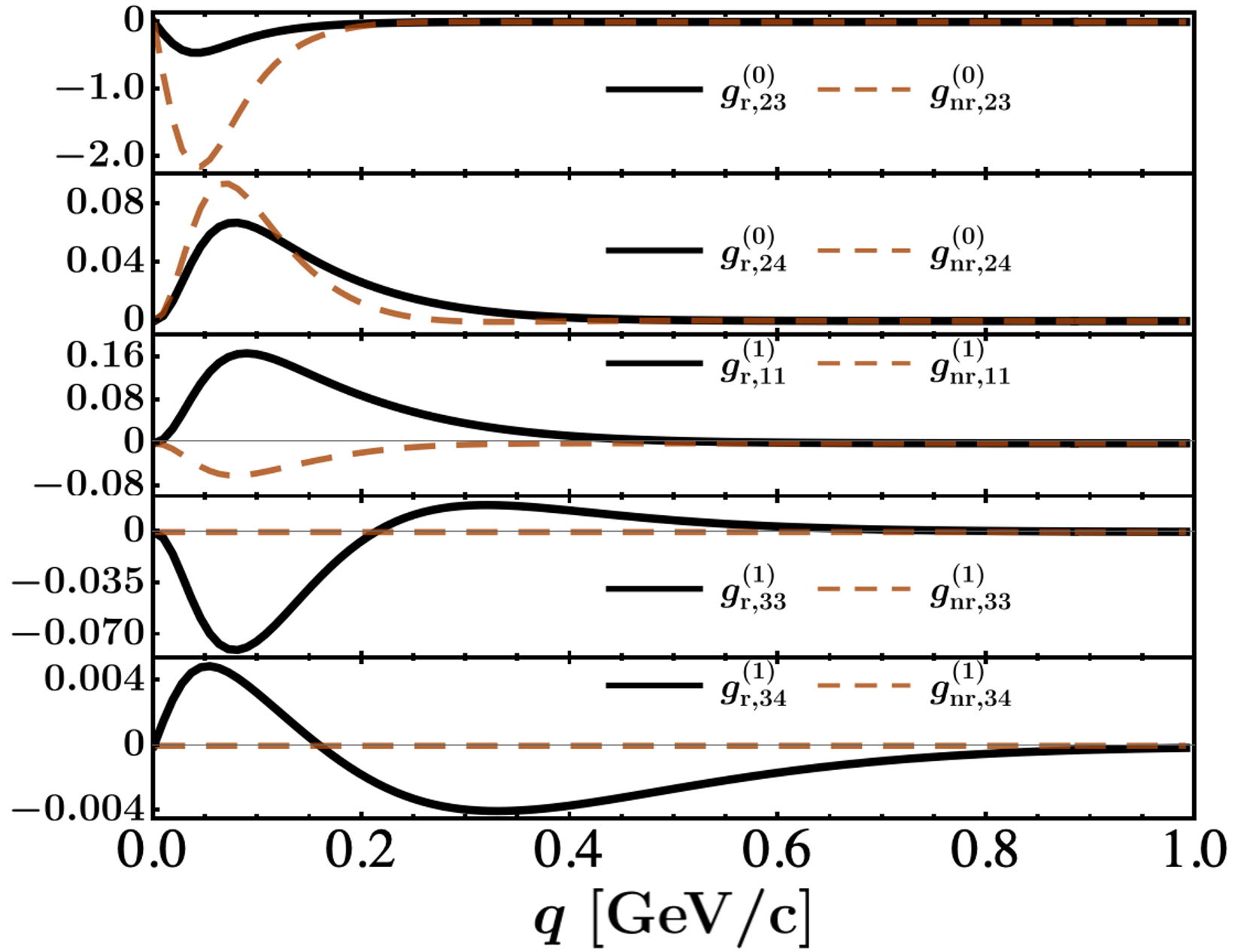}
  \caption{
    Plots of typical non-dominant components (dashed lines: non-relativistic components, solid lines: relativistic components) as functions of $q$. 
    The fixed arguments are $x_1=x_2=x_3=1/3$, $q_1 = q_2 = q_3 = q$, and $\langle \vec{q}_1, \vec{q}_2\rangle = \langle \vec{q}_1, \vec{q}_3\rangle = \langle \vec{q}_2, \vec{q}_3\rangle = 2\pi/3$.
    The component functions are in units of $\mathrm{GeV^{-3}}$.
  }
  \label{Fig:qqq120_2}
\end{figure}

For completeness, we also display the comparison for representative non-dominant components as functions of $q$ in Fig.~\ref{Fig:qqq120_2}.
The deviations of the non-dominant components appear in both the non-relativistic and relativistic domains.
Some of the non-relativistic components, such as $g_{\mathrm{nr},33}^{(1)}$ and $g_{\mathrm{nr},34}^{(1)}$, vanish identically, whereas their relativistic analogs remain finite.
Even more striking, the non-relativistic and relativistic components, $g_{\mathrm{nr},11}^{(1)}$ and $g_{\mathrm{r},11}^{(1)}$, carry opposite signs.
These contrasts underscore the role of relativistic dynamics in reshaping the wave functions.

\begin{figure*}[h!]
  \centering
  \includegraphics[width=0.45\linewidth]{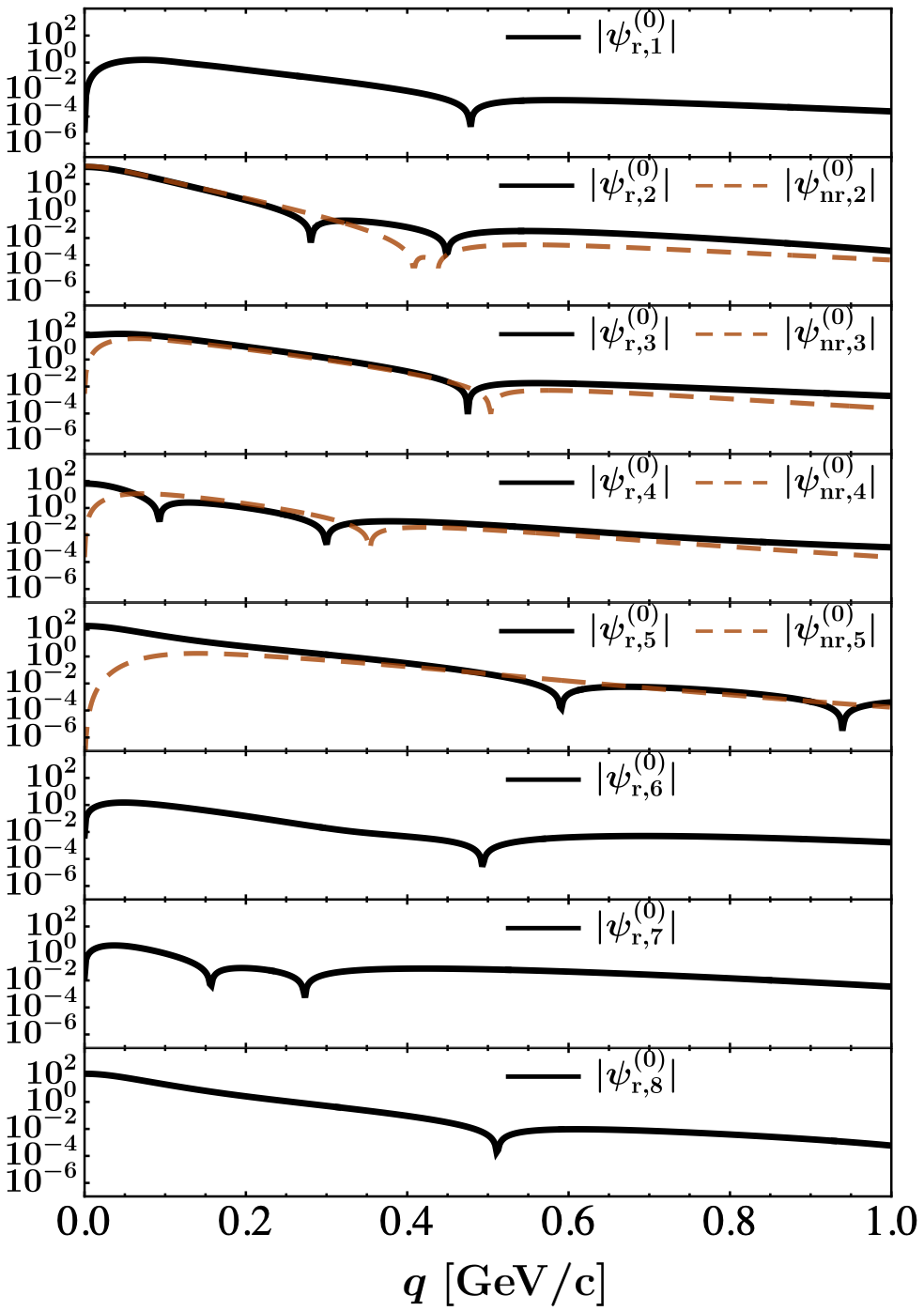}
  \includegraphics[width=0.45\linewidth]{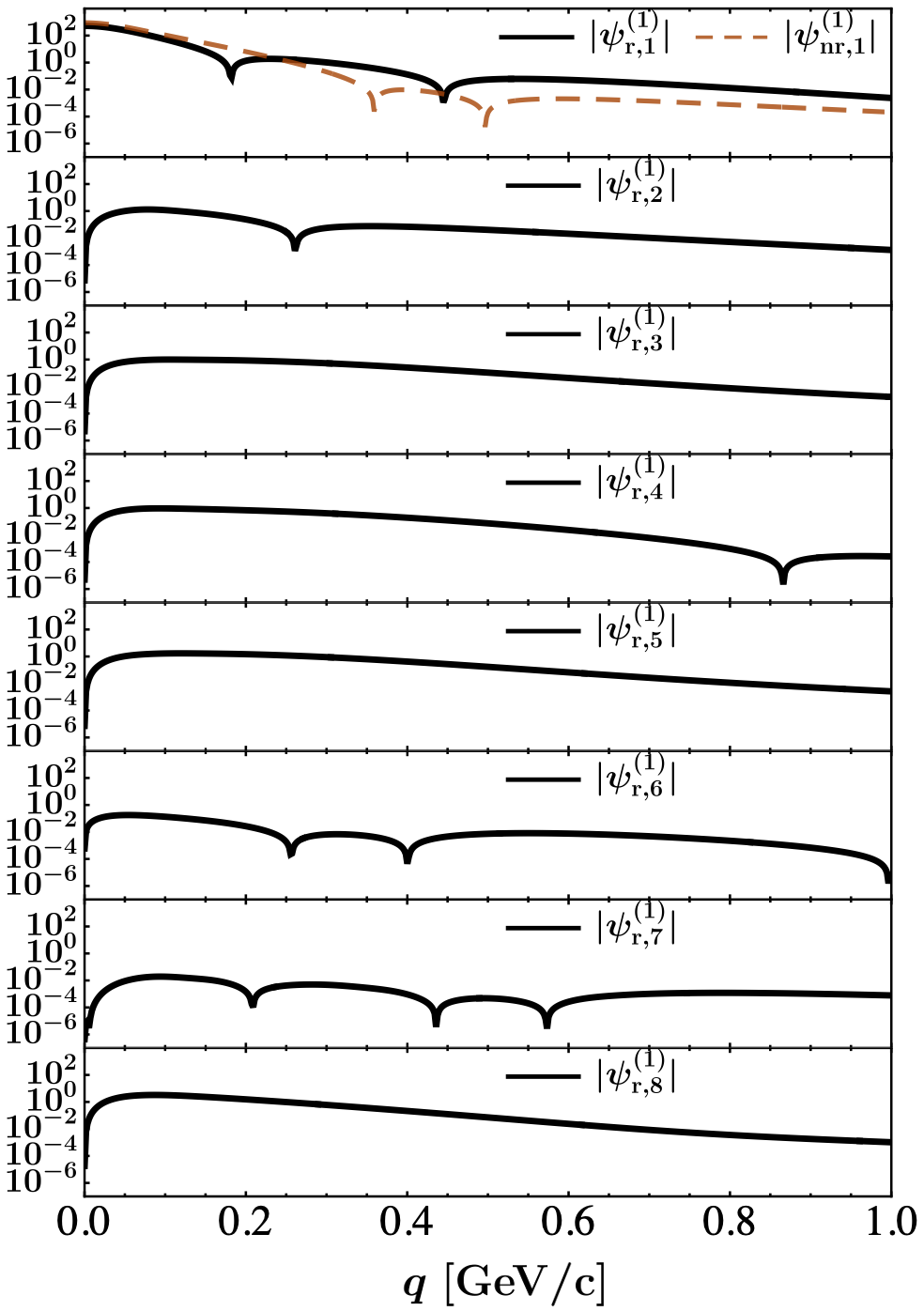}\\
  \includegraphics[width=0.45\linewidth]{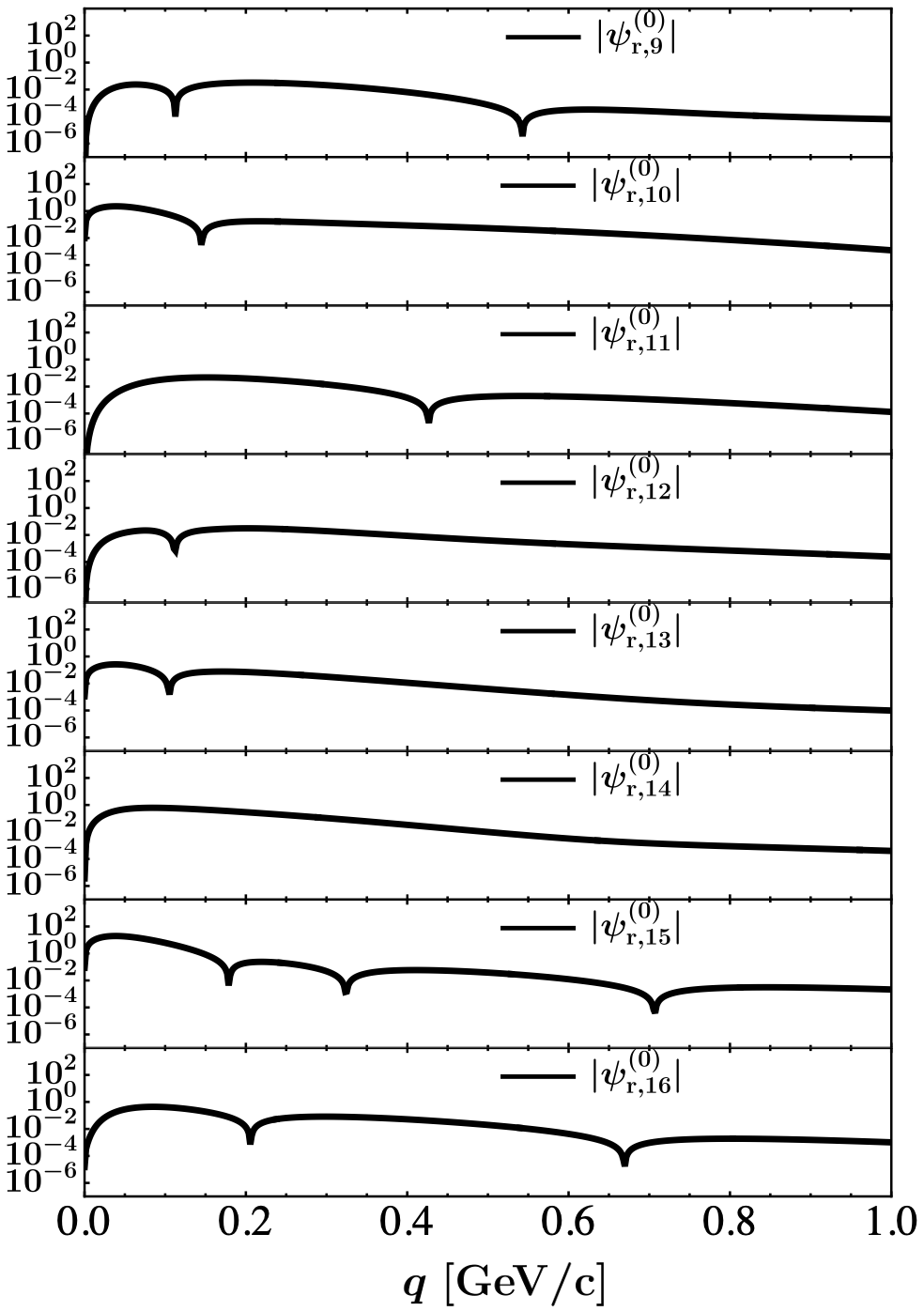}
  \includegraphics[width=0.45\linewidth]{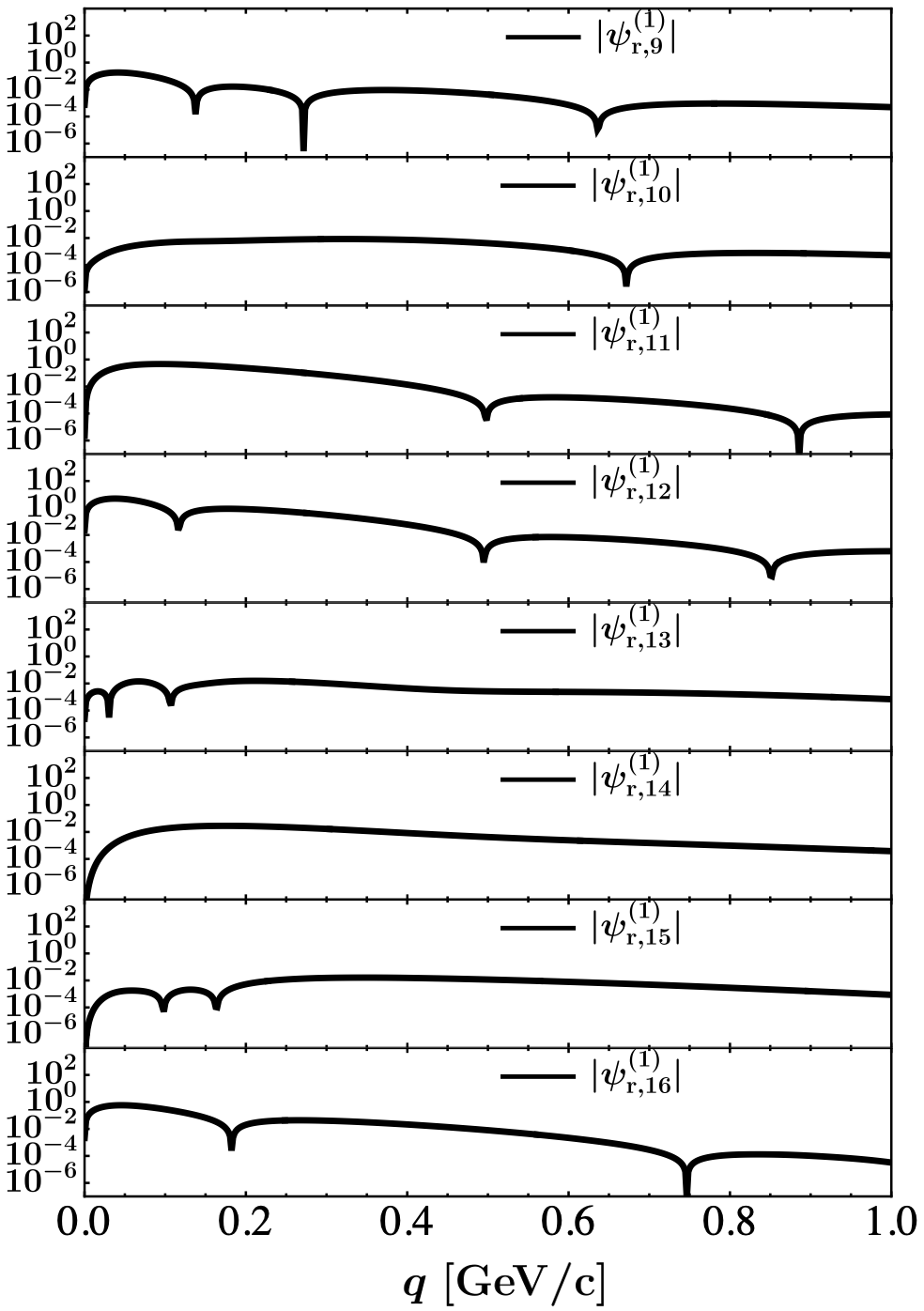}
  \caption{
    Logarithmic plots of dominant components (dashed lines: non-relativistic components, solid lines: relativistic components) as functions of $q$. 
    The fixed arguments are $x_1 = x_2 = x_3 = 1/3$, $\vec{q}_1 = (0, q, 0)$, $\vec{q}_2 = (q, 0, 0)$, $\vec{q}_3 = -\vec{q}_1 - \vec{q}_2$, and $\vec{n} = (0,0,1)$.
    All momenta are in units of GeV.
    The wave functions are in units of $\mathrm{GeV^{-3}}$.
  }
  \label{Fig:q_perp_q}
\end{figure*}




Comparison between the 32 relativistic wave functions, $\psi^{(t)}_{\mathrm{r},n}$ ($n=1 - 16,\;t=0,1$), and five non-relativistic ones, 
$\psi^{(t)}_{\mathrm{nr},n}$ ($n=1, t=1$ and $n=2-5, t=0$), is shown in  the Figure~\ref{Fig:q_perp_q}.
Some of the wave functions $\psi^t_{\mathrm{r},n}$ remain non-zero in the non-relativistic limit. 
In particular, the relativistic wave functions $\psi_{\mathrm{r},1}^{(1)}$ and $\psi_{\mathrm{r},2}^{(0)}$ are the dominant terms, just as in the non-relativistic solution, and in the low-momentum regime ($q\ll m$) they coincide almost exactly with their non-relativistic counterparts.
The next three wave functions $\psi_{\mathrm{r},3-5}^{(0)}$, which also have non-relativistic counterparts, show small but noticeable deviations even at low $q$.
Moreover, the purely relativistic components with no non-relativistic counterparts appear in the relativistic solution. Their amplitudes in the $q\ll m$ domain remain small, reflecting the comparatively weak but nonzero effect of relativistic dynamics, as expected.

\begin{figure}
  \centering
  \includegraphics[width=0.92\linewidth]{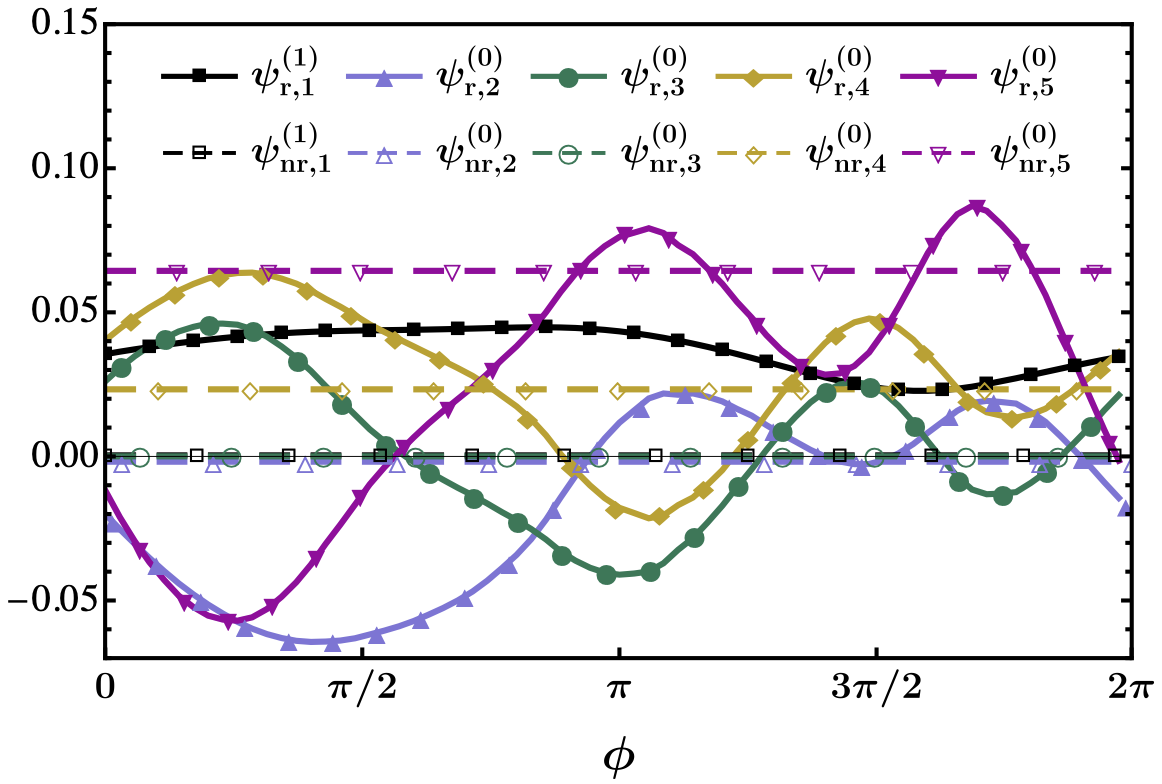}
  \caption{
    Plots of dominant components as functions of $\phi$.
    The fixed arguments are in Eqs.~(\ref{vec}-\ref{vec4}) with $\theta_2=\theta_3 =\pi/2$, $\phi_2=0$, $\phi_3=4\pi/7$, $\theta_n = \pi/7$, and $q=0.5$ GeV/c.
    The component functions are in units of $\mathrm{GeV^{-3}}$.  }
  \label{Fig:psi_phi05}
\end{figure}

\begin{figure}
  \centering
      \includegraphics[width=0.92\linewidth]{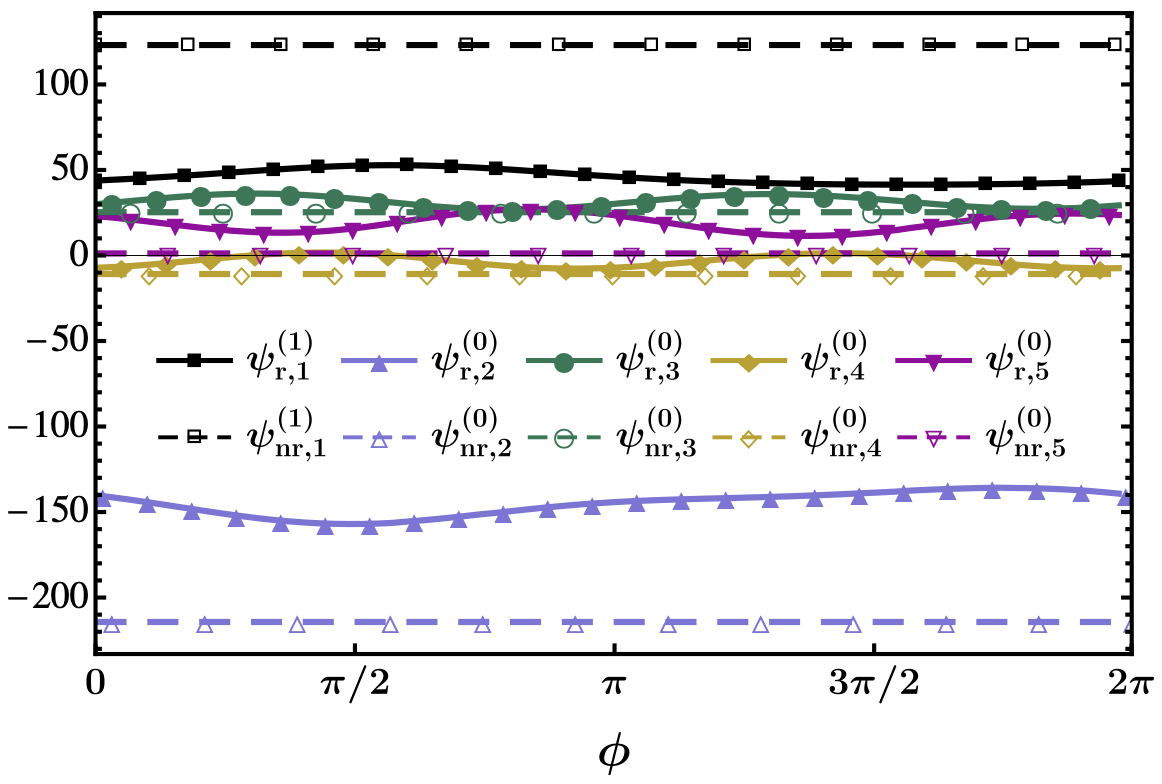}
  \caption{ The same as in Fig.  \protect{\ref{Fig:psi_phi05}} but for $q=0.1$ GeV/c.}
  \label{Fig:psi_phi01}
\end{figure}

\begin{figure}
\centering
\includegraphics[width=0.95\linewidth]{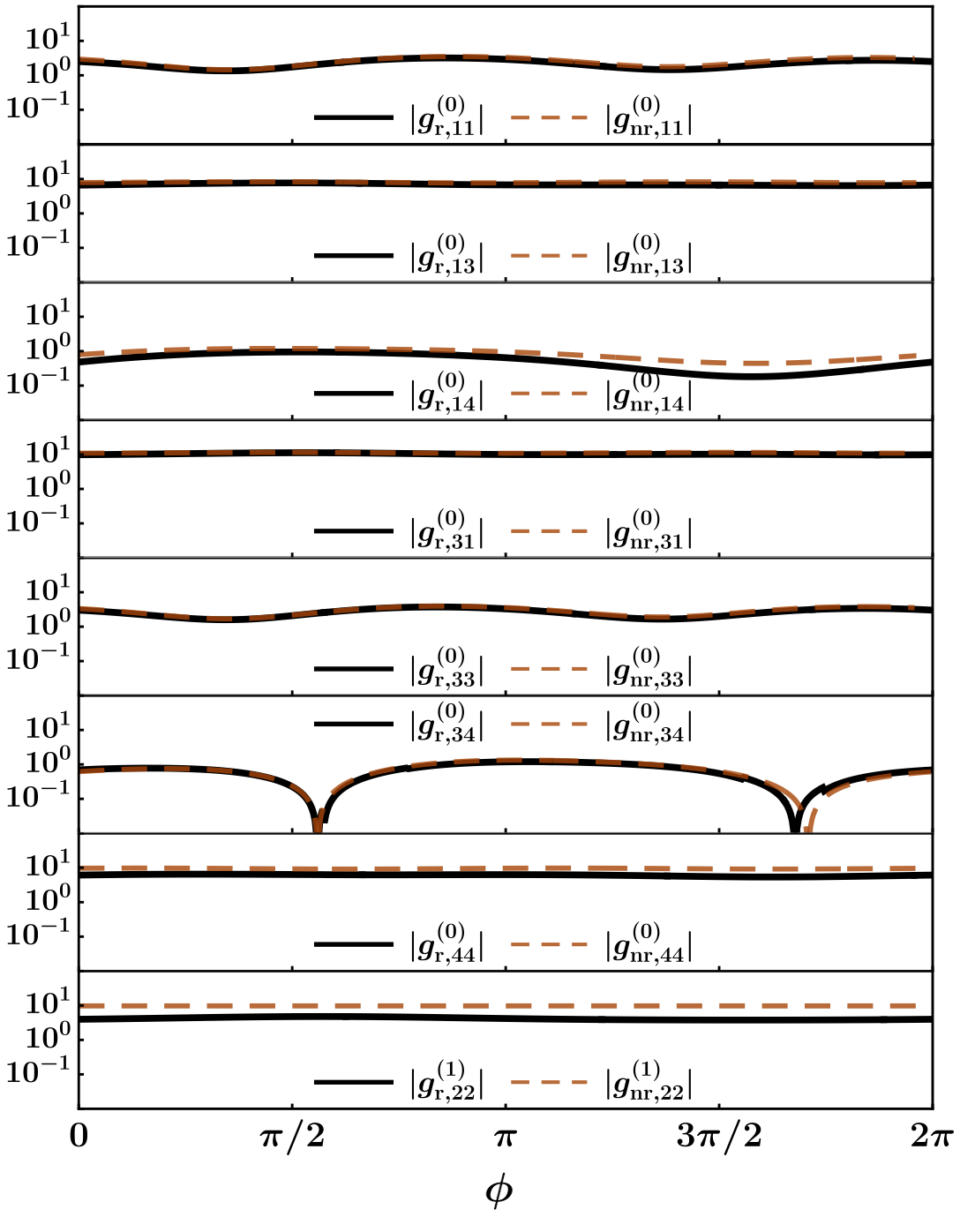}
\caption{
  Logarithmic plots of dominant components (dashed lines: non-relativistic components, solid lines: relativistic components) as functions of $\phi$. 
  The fixed arguments are in Eqs.~(\ref{vec}-\ref{vec4}) with $\theta_2=\theta_3 =\pi/2$, $\phi_2=0$, $\phi_3=2\pi/3$, $\theta_n = \pi/6$, and $q=0.1$ GeV/c.
  The component functions are in units of $\mathrm{GeV^{-3}}$.
}
\label{Fig:phi}
\end{figure}

\subsection{$\vec{n}$-dependence}\label{n}

The non-relativistic $^3$He wave function depends on three variables:  two independent momenta (for example, the Jacobi momenta) and the angle between them. 
The dependence on $\vec{n}$ adds two variables: two angles that define the orientation of the unit vector $\vec{n}$. 
For these two angles,
we take the angle $\theta_n$ between $\vec{n}$ and the perpendicular to the plane formed by the vectors $\vec{q}_{1,2,3}$ and the azimuthal angle $\phi$ in this plane. We will fix the value of $\theta_n$ and vary $\phi$. 
Since the $\vec{n}$-dependence is of relativistic origin, we should observe the $\phi$-dependence in the relativistic domain and its disappearance in the non-relativistic limit. The calculations in the present section are aimed to the test, whether our solution satisfy this expected observation.
 
The momenta $\vec{q}_i$ and vector $\vec{n}$ are
\begin{align}\label{vec}
  \vec{q}_1&=(q_{1x},q_{1y},q_{1z})
  \nonumber\\
  &=(-q_{2x}-q_{3x},-q_{2y}-q_{3y},-q_{2z}-q_{3z}),\\
  \vec{q}_2&=(q_{2x},q_{2y},q_{2z})
  \nonumber\\
  &=(q_2\sin\theta_2\cos\phi_2,q_2\sin\theta_2\sin\phi_2,q_2\cos\theta_2),\\
  \vec{q}_3&=(q_{3x},q_{3y},q_{3z})
  \nonumber\\
  &=(q_3\sin\theta_3\cos\phi_3,q_3\sin\theta_3\sin\phi_3,q_3\cos\theta_3),\\
  \vec{n}&=(n_x,n_y,n_z)\nonumber\\
  &=(\sin\theta_n\cos\phi,\sin\theta_n\sin\phi,\cos\theta_n).\label{vec4}
\end{align}
We take $q_2=q_3=q$, and fix $\theta_2=\theta_3=\pi/2$, such that both 
$\vec{q}_{2,3}$ lie in the $xy$-plane. 
We set $\phi_2=0$ placing $\vec{q}_2$ directly along $x$-axis, and $\phi_3=\frac{4}{7}\pi$ placing $\vec{q}_3$ in the second quadrant, 
where the choice is to keep the transform matrix $M_{ij,n}$ invertible.
We choose $\theta_n=\frac{1}{7}\pi$, while the azimuthal angle $\phi$ of the vector $\vec{n}$ remain unfixed.

In Appendix~\ref{app1}, the transverse vectors are
\begin{equation}\label{kt}
  \vec{R}_{i\perp}=\vec{q}_i-(\vec{n}\cd\vec{q}_i)\vec{n}.
\end{equation}
By Eq.~(\ref{kt}) we construct three transverse vectors $\vec{R}_{1\perp}$, $\vec{R}_{2\perp}$ and $\vec{R}_{3\perp}$.
The modules of transversed components are $R_{i\perp}=\sqrt{\vec{R}_{i\perp}^{\,2}}=\sqrt{\vec{q}^{\,2}_i-(\vec{n}\cd\vec{q}_i)^2}$.
The variables $x_i$ for $i=1,2,3$ are
\begin{equation}\label{xi}
  x_i=\frac{\varepsilon_{q_i}-\vec{n}\cd\vec{q}_i}{\varepsilon_{q_1}+\varepsilon_{q_2}+\varepsilon_{q_3}},\quad \varepsilon_{q_i}=\sqrt{m^2+\vec{q}^{\,2}_i}.
\end{equation}
These variables satisfy the conditions:
\begin{align*}
  &\vec{q}_1+\vec{q}_2+\vec{q}_3=0,\quad \vec{k}_{1\perp}+\vec{k}_{2\perp}+\vec{k}_{3\perp}=0,
 \\ 
   &x_1+x_2+x_3=1.
\end{align*}

Substituting the expressions for $\vec{R}_{i\perp},x_i$ in terms of $\vec{q}_i$, Eqs.~(\ref{vec}-\ref{vec4}), as arguments of the wave function, 
we analyze the dependence on $\phi$,at first, in the relativistic domain ($q = 0.5$ GeV/c). 

Figure~\ref{Fig:psi_phi05} compares the five relativistic wave functions $\psi_{1-5}$ at the 
momentum $q=0.5$ GeV/c with extrapolations of their non-relativistic counterparts as functions of the angle $\phi$, defined in Eq.~(\ref{vec4}).
All the relativistic wave functions 
exhibit strong $\vec{n}$ dependence, that is, on relative orientation of the LF plane and the particle momenta, as expected.

Figure~\ref{Fig:psi_phi01} shows the same comparison as in Figure~\ref{Fig:psi_phi05}  but for smaller momentum $q=0.1$ GeV/c.
The relativistic wave functions with zero angular momentum, $\psi_{\mathrm{r},1}^{(0)}$ and $\psi_{\mathrm{r},2}^{(0)}$, exhibit weak $\vec{n}$ dependence in the non-relativistic limit.
The relativistic wave functions with nonzero angular momentum, $\psi_{\mathrm{r},3-5}^{(0)}$, display a noticeable $\vec{n}$ dependence which, however, is much weaker than in Figure~\ref{Fig:psi_phi05}, where momentum $q$ is five times larger.
Besides, their magnitudes are small and therefore they don’t make crucial influence on the total momentum distribution of $^3$He.

Similar comparison as in Figure~\ref{Fig:psi_phi01}  but in terms of the functions $g_{ij}^{(t)}$ is shown in 
Figure~\ref{Fig:phi}. It presents eight dominant scalar functions $g_{ij}^{(t)}$, seven components for $t=0$ and one component for $t=1$, as functions of $\phi$ at fixed $q=0.1$ GeV/c.
Each of these dominant components
has a magnitude of the order $1$ GeV$^{-3}$.
The comparison shows that in the non-relativistic domain (at small $q\ll m$),  the relativistic components align closely with their $\phi$-independent non-relativistic counterparts.

The results shown in Figures~\ref{Fig:psi_phi01}  and \ref{Fig:phi} demonstrate that in the  non-relativistic limit $q\ll m$ our relativistic $^3$He wave function, defined on the LF plane, stops to depend on orientation of this plane (in contrast to the relativistic domain $q\sim 0.5$ GeV/c, Figure~~\ref{Fig:psi_phi05}), as should be for any non-relativistic wave function. This confirms validity of our results. General reason of this property is in the fact that the non-relativistic limit can be formally achieved, taking the limit $c\to\infty$ for the speed of light $c$. Then the equation for the LF plane (with restored speed of light): 
$\omega_0t-\frac{1}{c}\vec{\omega}\cd\vec{r}=0$ turns  into $t=0$, that is, LFD turns into the instant form of dynamics, without any dependence on orientation 
of the LF plane, that is, on the angles determining orientation of $\vec{\omega}=|\vec{\omega}|\vec{n}$.

\subsection{$x_{12}$-dependence}\label{x12_dependence}
\begin{figure}
  \centering
  \includegraphics[width=0.92\linewidth]{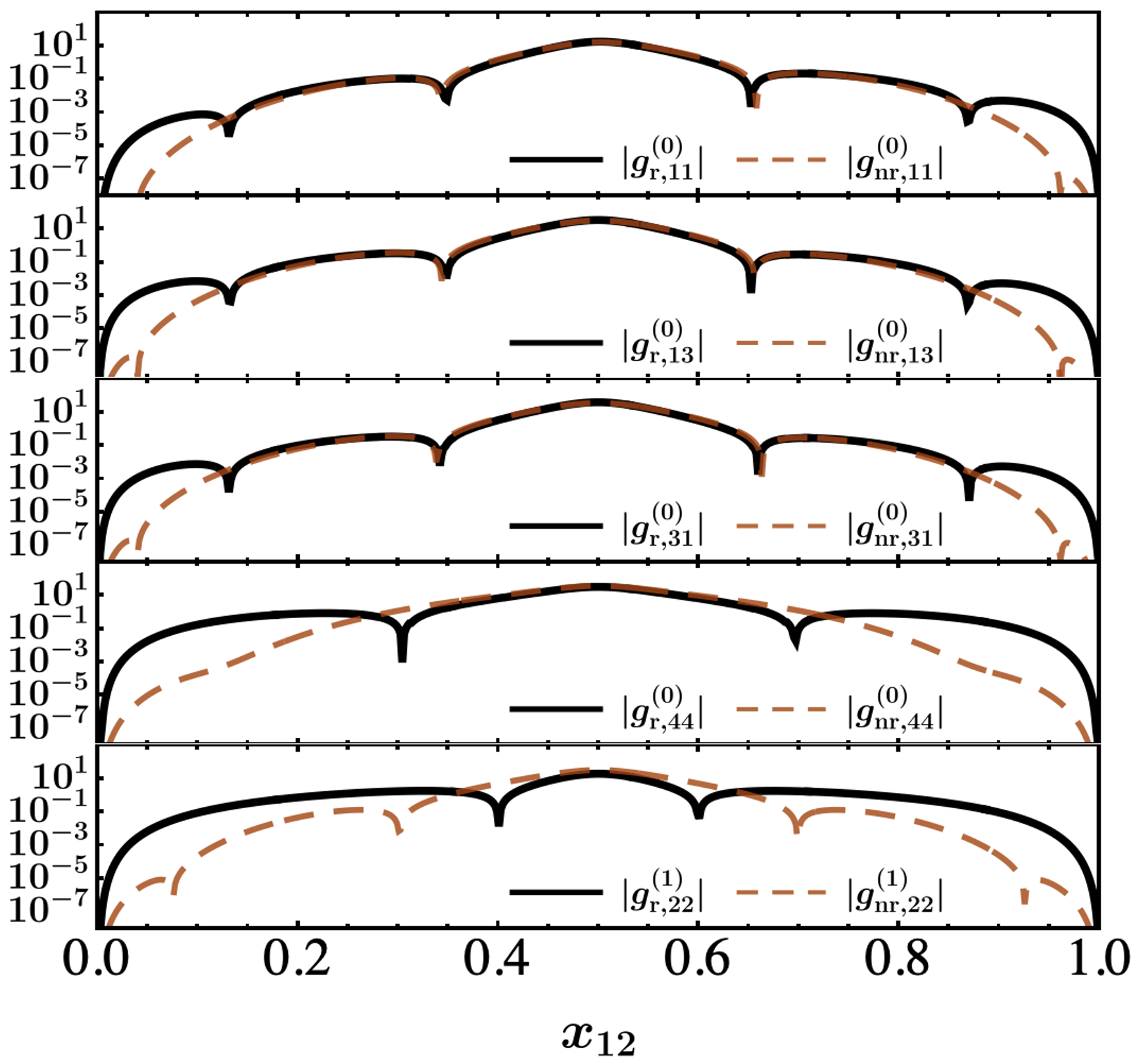}
  \caption{
    Logarithmic plots of dominant components (dashed lines: non-relativistic components, solid lines: relativistic components) as functions of $x_{12}$. 
    The fixed arguments are $x_{1} = x_{12}(1-x_3)$, $x_3=1/3$, $x_2 = 1 -x_1-x_3$, $\vec{R}_{1\perp} = (0, 0.05)$, $\vec{R}_{2\perp} = (0.05,0)$, and $\vec{R}_{3\perp} = -\vec{R}_{1\perp}-\vec{R}_{2\perp}$. 
    All momenta are in units of GeV/c.
    The component functions are in units of $\mathrm{GeV^{-3}}$.
  }
  \label{Fig:x12}
\end{figure}

The dependence on the light-front momentum fractions is also illuminating.
In Fig.~\ref{Fig:x12}, we present five dominant scalar functions $g_{ij}^{(t)}$, four components for $t=0$ and one component for $t=1$ as functions of the ratio $x_{12}$, where $x_{12} = \frac{x_1}{1-x_3}$.
At our chosen kinematics, each dominant component has a magnitude of order $10\;\mathrm{GeV}^{-3}$.
Near the non-relativistic regime (around $x_{12}=1/2$), the relativistic components overlap their non-relativistic counterparts almost exactly.  
Similar to Figs.~\ref{Fig:qqq120} and \ref{Fig:phi}, the deviations between the two components only become significant beyond the non-relativistic regime, reflecting the relativistic corrections to the wave functions in the longitudinal momentum fraction degree of freedom.

\section{Conclusion}\label{concl}

We have calculated the relativistic LF $^3$He wave function. 
In this calculation, we used the explicitly covariant version of LFD \cite{cdkm}. 
The main advantage of covariance is two-fold: 1) it ensures that the total angular momentum of the wave function is definite and equal to 1/2; 2) it allows us to decompose the wave function into covariant spin-isospin structures with the scalar coefficients (spin components). 
The total number of these components is 32, each of them depends on five scalar variables, and a full set of them completely determines the LF $^3$He wave function. 
The spin-isosipin components satisfy the system of equations, which we solved iteratively. 
For interaction, we have taken a one-boson exchange kernel determined by the exchange of seven mesons (Table \ref{tab3}) with the parameters (masses, coupling constants, vertex form factors) found in~\cite{M1}. 
This kernel is used “as is,” \ie, in its original relativistic form, without making the potential approximation.

Using the $^3$He wave function found in this framework, one can calculate the $^3$He electromagnetic form factors at large momentum transfer 
and compare them with existing \cite{Jlab_He3} and forthcoming \cite{JLab} experimental data.

The aim of this publication is not only to calculate the relativistic LF $^3$He wave function, but to demonstrate how the LF wave function for a system of a few fermions can be found at all. 
In particular,this approach provides a direct path to the nucleon LF wave function, whose Fock-space decomposition contains not only the three-quark sector but also higher components, including gluons.
The structure of the LF wave function corresponding to the three-quark sector is similar to the $^3$He wave function. As mentioned, in the standard version of LFD, without constructing covariant spin structures and calculating the invariant coefficients at the front of them, the nucleon LFWF was found  numerically in \cite{BLFQ}. The analytical methods developed in the present work reduce the amount of numerical calculations.

In the same spirit, this method can be extended to hypernuclei such as $^3_\Lambda H$ and $^3_\Lambda He$, as well as to heavier nuclei, including $^4$He, {\it etc.}
Technical difficulties, of course, increase when more constituents are involved, but the principal constructions and ideas remain the same as in the present work.

\section*{Acknowledgement}

VAK was supported by the Chinese Academy of Sciences President's International Fellowship Initiative (PIFI), Grant No. 2023VMA0010, during his visits to Institute of Modern Physics of Chinese Academy of Sciences, where part of this work was fulfilled.
VAK is sincerely grateful to the light-front QCD group in IMP for kind hospitality during his visits.

Z. Zhu is supported by China Association for Science and Technology, and was supported by the Natural Science Foundation of Gansu Province China Grant No. 23JRRA571. Z. Zhu and Z. Zhang gratefully acknowledge financial support from China Scholarship Council.
This research is supported by Gansu International Collaboration and Talents Recruitment Base of Particle Physics (2023–2027), by the Senior Scientist Program funded by Gansu Province Grant No. 25RCKA008.
\bigskip
\appendix

\section{Matrices and coefficients determining the basis (\protect{\ref{chi_n}})}\label{appendix1}

The basis functions $\chi_n$ are represented in the short covariant form by Eq.~(\ref{chi_n}). 
They are expressed in terms of matrices $O$ and coefficients $c_n$ defined below.
In these formulas, $q$ and $k$ are the Jacobi 4-momenta given in Eq.~(\ref{Jacpq}). 
$|\vec{q}|$ and $|\vec{k}|$ are the modules of the spatial parts of these 4-vectors in the frame where $\vec{\cal P}=0$ with ${\cal P}$ defined in Eq.~(\ref{calP}), that is, $\varepsilon_q=\frac{{\cal P}\cd q}{\cal M}$ with ${\cal M}^2={\cal P}^2$, and similarly for 
$|\vec{k}|$. 
$\vec{k}\cd\vec{q}=\frac{1}{{\cal M}^2}({\cal P}\cd k)({\cal P}\cd q)- k\cd q$.
The matrices $O$ and coefficients $c_n$ of each $\chi_n$ in Eq.~(\ref{chi_n}) are
\begin{widetext}
  \begin{align}\label{OO}
    &c_1=-\frac{i}{4\pi\sqrt{2}}c, 
    && O_{1}^{(A1)} = \Pi_+ \gamma_5,
    &&&& O_{1}^{(B1)} = \Pi_+;
 \\
    &c_2=-\frac{i}{4\pi\sqrt{6}}c, 
    && O_{2}^{(A1)} = \Pi_+\gamma_{\mu}\Pi_-,
    &&&& O_{2}^{(B1)} = \Pi_+\gamma_{\mu}\gamma_5\Pi_+ ;
 \\
    &c_3=-\frac{i\sqrt{3}}{8\pi}c, 
    && O_{3}^{(A1)} =\frac{1}{3} \Pi_+\gamma_{\mu}\Pi_- =\frac{1}{3}O_2^{(A1)},
    &&&& O_{3}^{(B1)} = \Pi_+\gamma_{\mu}\gamma_5 \Pi_+=O_{2}^{(B1)} ,\nonumber
    \\
    &
    &&O_{3}^{(A2)} = \frac{1}{|\vec{k}|}\Pi_+\hat{k}\Pi_-,
    &&&&O_{3}^{(B2)} = \frac{1}{|\vec{k}|} \Pi_+\hat{k}\gamma_5\Pi_+;
   \\
    &c_4=-\frac{i\sqrt{3}}{8\pi}c, 
    &&O_{4}^{(A1)}=\frac{1}{3} \Pi_+\gamma_{\mu}\Pi_-=\frac{1}{3}O_2^{(A1)}, 
    &&&&O_{4}^{(B1)} = \Pi_+\gamma_{\mu}\gamma_5 \Pi_+=O_{2}^{(B1)},\nonumber
    \\
    &
    &&O_{4}^{(A2)} = \frac{1}{|\vec{q}|} \Pi_+\hat{q}\Pi_-,
    &&&&O_{4}^{(B2)} = \frac{1}{|\vec{q}|} \Pi_+\hat{q}\gamma_{5}\Pi_+;
  \\
    &c_5=-\frac{3i\sqrt{6}}{16\pi}c,
    &&O_{5}^{(A1)}= \frac{1}{9}\Pi_+\gamma_{\mu}\Pi_-=\frac{1}{9}O_2^{(A1)},
    &&&&O_{5}^{(B1)}= \Pi_+\gamma_{\mu}\gamma_5 \Pi_+=O_{2}^{(B1)},
   \nonumber \\
    &
    &&O_{5}^{(A2)}= \frac{1}{3|\vec{k}|} \Pi_+\hat{k}\Pi_-=\frac{1}{3}O_{3}^{(A2)},
    &&&&O_{5}^{(B2)} =  \frac{1}{|\vec{k}|} \Pi_+\hat{k} \gamma_{5}\Pi_+=O_{3}^{(B2)},\nonumber
    \\
    &
    &&O_{5}^{(A3)} = \frac{1}{3|\vec{q}|} \Pi_+\hat{q}\Pi_- =\frac{1}{3}O_{4}^{(A2)} ,
    &&&&O_{5}^{(B3)} =  \frac{1}{|\vec{q}|} \Pi_+\hat{q}\gamma_{5}\Pi_+=O_{4}^{(B2)},\nonumber
    \\      
    &
    &&O_{5}^{(A4)}=- \frac{\vec{k}\cd\vec{q}}{|\vec{q}||\vec{k}|}\frac{1}{|\vec{k}|}\Pi_+\hat{k}\Pi_-=
    - \frac{\vec{k}\cd\vec{q}}{|\vec{q}||\vec{k}|}O_{3}^{(A2)},
    &&&&O_{5}^{(B4)} = \frac{1}{|\vec{q}|}\Pi_+\hat{q}\gamma_{5}\Pi_+=O_{4}^{(B2)};
  \\ 
    &c_6=-\frac{3i}{\sqrt{10}}\frac{1}{8\pi}c,
    &&O_6^{(A1)}=\frac{2}{3}\frac{\vec{k}\cd \vec{n}}{|\vec{k}|}\Pi_+\gamma^{\mu}\Pi_-=\frac{2}{3}\frac{\vec{k}\cd \vec{n}}{|\vec{k}|}O_2^{(A1)},
    &&&&O_6^{(B1)}=\Pi_+ \gamma_{\mu}\gamma_5\Pi_+=O_2^{(B1)},\nonumber
    \\
    &
    &&O_6^{(A2)}=\frac{{\cal M}}{(\omega\cd p)} \Pi_+\hat{\omega}\Pi_-,
    &&&&O_6^{(B2)}=\frac{1}{|\vec{k}|}\Pi_+\hat{k}\gamma_5\Pi_+=O_3^{(B2)},\nonumber
    \\
    &
    &&O_6^{(A3)}=  \frac{1}{|\vec{k}|}\Pi_+\hat{k}\Pi_-=O_3^{(A2)},
    &&&&O_6^{(B3)}=\frac{{\cal M}}{(\omega\cd p)}\Pi_+\hat{\omega}\gamma_5\Pi_+;
   \\
    &c_7=-\frac{3i}{\sqrt{10}}\frac{1}{8\pi}c,
    &&O_7^{(A1)}=\frac{2}{3}\frac{\vec{q}\cd \vec{n}}{|\vec{q}|}\Pi_+\gamma^{\mu}\Pi_-=\frac{2}{3}\frac{\vec{q}\cd \vec{n}}{|\vec{q}|}O_2^{(A1)},
    &&&&O_7^{(B1)}=\Pi_+ \gamma_{\mu}\gamma_5\Pi_+=O_2^{(B1)},\nonumber
    \\
    &
    &&O_7^{(A2)}=\frac{{\cal M}}{(\omega\cd p)} \Pi_+\hat{\omega}\Pi_-=O_6^{(A2)},
    &&&&O_7^{(B2)}=\frac{1}{|\vec{q}|}\Pi_+\hat{q}\gamma_5\Pi_+=O_4^{(B2)},\nonumber
    \\
    &
    &&O_7^{(A3)}=  \frac{1}{|\vec{q}|} \Pi_+\hat{q}\Pi_-=O_4^{(A2)},
    &&&&O_7^{(B3)}=\frac{{\cal M}}{(\omega\cd p)}\Pi_+\hat{\omega}\gamma_5\Pi_+=O_6^{(B3)};
    \\
    &c_8=-\sqrt{\frac{3}{2}}\frac{i}{8\pi} c,
    &&O_8^{(A1)}=\frac{1}{|\vec{k}|} \Pi_+\hat{k}\Pi_-=O_3^{(A2)},
    &&&&O_8^{(B1)}=\frac{1}{|\vec{q}|}\Pi_+\hat{q}\gamma_5\Pi_+=O_4^{(B2)},\nonumber
    \\
    &
    &&O_8^{(A2)}=-\frac{1}{|\vec{q}|} \Pi_+\hat{q}\Pi_-=-O_4^{(A2)},
    &&&&O_8^{(B2)}=\frac{1}{|\vec{k}|}\Pi_+\hat{k}\gamma_5\Pi_+=O_3^{(B2)};
  \\
    &c_9=-\sqrt{\frac{3}{2}}\frac{i}{8\pi} c,
    &&O_9^{(A1)}=\frac{1}{|\vec{k}|} \Pi_+\hat{k}\Pi_-=O_3^{(A2)},
    &&&&O_9^{(B1)}=\frac{\cal M}{(\omega\cd p)}\Pi_+\hat{\omega}\gamma_5\Pi_+=O_6^{(B3)},\nonumber
    \\
    &
    &&O_9^{(A2)}=-\frac{\cal M}{(\omega\cd p)}  \Pi_+\hat{\omega}\Pi_- =-O_6^{(A2)},
    &&&&O_9^{(B2)}=\frac{1}{|\vec{k}|} \Pi_+\hat{k}\gamma_5\Pi_+ =O_3^{(B2)};
    \\
    &c_{10}=-\sqrt{\frac{3}{2}}\frac{i}{8\pi} c,
    &&O_{10}^{(A1)}=\frac{1}{|\vec{q}|}  \Pi_+\hat{q}\Pi_- =O_4^{(A2)},
    &&&&O_{10}^{(B1)}=\frac{\cal M}{(\omega\cd p)} \Pi_+\hat{\omega}\gamma_5\Pi_+ =O_6^{(B3)},\nonumber
    \\
    &
    &&O_{10}^{(A2)}=-\frac{\cal M}{(\omega\cd p)}  \Pi_+\hat{\omega}\Pi_- =-O_6^{(A2)},
    &&&&O_{10}^{(B2)}=\frac{1}{|\vec{q}|} \Pi_+\hat{q}\gamma_5\Pi_+ =O_4^{(B2)};
   \\
    &c_{11}=-\frac{\sqrt{3}}{8\pi}c,
    &&O_{11}^{(A1)}=\frac{1}{|\vec{k}||\vec{q}|{\cal M}}\epsilon_{\mu\nu\alpha\beta}{\cal P}_{\mu}k_{\nu}q_{\alpha}\; \Pi_+\gamma^{\beta}
    \Pi_-  ,
    &&&&O_{11}^{(B1)}= \Pi_+  =O_1^{(B1)};
    \\
    &c_{12}=-\frac{\sqrt{3}}{8\pi}c,
    &&O_{12}^{(A1)}=\frac{1}{|\vec{k}|({\cal P}\cd \omega)}\epsilon_{\mu\nu\alpha\beta}{\cal P}_{\mu}k_{\nu}\omega_{\alpha}\; \Pi_+\gamma^{\beta}
    \Pi_-  ,
    &&&&O_{12}^{(B1)}= \Pi_+  =O_1^{(B1)};
    \\
    &c_{13}=-\frac{\sqrt{3}}{8\pi}c,
    &&O_{13}^{(A1)}=\frac{1}{|\vec{q}|({\cal P}\cd \omega)}\epsilon_{\mu\nu\alpha\beta}{\cal P}_{\mu}q_{\nu}\omega_{\alpha}\; \Pi_+\gamma^{\beta}
    \Pi_-  ,
    &&&&O_{13}^{(B1)}= \Pi_+  =O_1^{(B1)};
   \\
    &c_{14}=-\frac{\sqrt{3}}{8\pi}c,
    &&O_{14}^{(A1)}=\frac{1}{|\vec{k}||\vec{q}|{\cal M}}\epsilon_{\beta\nu\alpha\mu}{\cal P}_{\beta}k_{\nu}q_{\alpha}\;
    \Pi_+\gamma_{5}\Pi_- ,
    &&&&O_{14}^{(B2)}= \Pi_+\gamma_{\mu}\gamma_5\Pi_+  =O_2^{(B1)};
   \\
    &c_{15}=-\frac{\sqrt{3}}{8\pi}c,
    &&O_{15}^{(A1)}=\frac{1}{|\vec{k}|(\omega\cd p)}\epsilon_{\beta\nu\alpha\mu}{\cal P}_{\beta}k_{\nu}\omega_{\alpha}
    \; \Pi_+\gamma_{5}\Pi_- ,
    &&&&O_{15}^{(B2)}= \Pi_+\gamma_{\mu}\gamma_5\Pi_+  =O_2^{(B1)};
  \\
    &c_{16}=-\frac{\sqrt{3}}{8\pi}c,
    &&O_{16}^{(A1)}=\frac{1}{|\vec{q}|(\omega\cd p)}\epsilon_{\beta\nu\alpha\mu}{\cal P}_{\beta}q_{\nu}\omega_{\alpha}
    \; \Pi_+\gamma_{5}\Pi_- ,
    &&&&O_{16}^{(B2)}= \Pi_+\gamma_{\mu}\gamma_5\Pi_+  =O_2^{(B1)}.
  \end{align}
\end{widetext}

\section{LF variables {$\vec{R}_{i\perp}$ and $x_i$}} \label{app1}
In the explicitly covariant version of LFD, the three-body wave function depends on the four-momenta $k_1$, $k_2$, $k_3$, $\omega\tau$, and $p$ (see Eq.~(\ref{Psi})).  
As mentioned, all the four-momenta lie on their respective mass shells $k_1^2=k_2^2=k_3^2=m^2$, $p^2=M^2$, and $(\omega\tau)^2=0$. They also satisfy the conservation law $k_1+k_2+k_3=p+\omega\tau$~\cite{cdkm}. 

From these four-momenta we construct the following ones:
\begin{eqnarray}
  R_1&=&k_1-x_1 p,\label{R123}
  \\
  R_2&=&k_2-x_2 p,
  \\
  R_3&=&k_3-x_3 p,\label{R123_3}
\end{eqnarray}
where $x_i=(\omega\cd k_i)/(\omega\cd p)$. 
These variables satisfy the conditions $R_i\cd\omega=0$. 
We represent $R_i$ in terms of its components: 
$R_i=(\vec{R}_{i\parallel},\vec{R}_{i\perp},R_{i0})$, where $\vec{R}_{i\parallel}\parallel \vec{\omega}$, $\vec{R}_{i\perp}\perp\vec{\omega}$, that is, 
$\vec{R}_{i\parallel}\cd\vec{\omega}=|\vec{R}_{i\parallel}||\vec{\omega}|=|\vec{R}_{i\parallel}|\omega_0$ and $\vec{R}_{i\perp}\cd\vec{\omega}=0$. From the condition $R_i\cd\omega=0$, it follows $|\vec{R}_{i\parallel}|=R_{i0}$, so $\vec{R}^2_{i\perp}=-R_i^2$ is Lorentz invariant, as are  the scalar products $\vec{R}_{i\perp}\cd\vec{R}_{j\perp}$.

The three-body LF wave function depends on the variables $\vec{R}_{1\perp},\vec{R}_{2\perp},\vec{R}_{3\perp}$ and $x_1,x_2,x_3$. Because of the momentum conservation,  they are not independent, but satisfy the relations: $\vec{R}_{1\perp}+\vec{R}_{2\perp}+\vec{R}_{3\perp}=0$, and $x_1+x_2+x_3=1$. 
By excluding, for example, $\vec{R}_{3\perp}$ and $x_3$, the scalar wave function, which is invariant under rotations and Lorentz transformations, reads:
\begin{align*}
  g_{ij}^t(1,2,3)&=g_{ij}^t(\vec{R}_{1\perp},\vec{R}_{2\perp},\vec{R}_{3\perp};x_1,x_2,x_3)
  \\
  &=g_{ij}^t(\vec{R}^2_{1\perp},\vec{R}^2_{2\perp},\vec{R}_{1\perp}\cd\vec{R}_{2\perp};x_1,x_2).
\end{align*}
That is, it depends on five scalar variables.

Let us prove  Eq.~(\ref{calM}).  We substitute the definition (\ref{R123}-\ref{R123_3}) of $R_i$ into the identity $\vec{R}^2_{i\perp}=-R_i^2$, and  obtain: 
\begin{align}\label{B2}
  \sum_{i=1}\frac{\vec{R}_{i\perp}^2+m_i^2}{x_i}&=\sum_{i=1}\frac{-(k_i-x_i p )^2+m_i^2}{x_i}
  \nonumber\\
  &=2p\cd\sum_{i=1}^n k_i -M^2
\end{align}
Squaring the equality $\sum_i k_i-p=\omega\tau$, we get:
$$
\Bigl(\sum_i k_i-p\Bigr)^2=\Bigl(\sum_i k_i\Bigr)^2-2p\cdot\sum_i k_i+M^2=(\omega\tau)^2=0.
$$
That  is,
$
2p\cdot\sum_i k_i-M^2=(\sum_i k_i)^2.
$
Substituting this equality into Eq.~(\ref{B2}), we obtain Eq.~(\ref{calM}).

\section{Kinematical relations}\label{kinemat}
The system of equations (\ref{gij}) is written for the functions $g^{(t)}_{ij}=g^{(t)}_{ij}(\vec{R}_{1\perp},\vec{R}_{2\perp},\vec{R}_{3\perp};x_1,x_2,x_3)$, where
$\vec{R}_{1\perp}+\vec{R}_{2\perp}+\vec{R}_{3\perp}=0$, and $x_1+x_2+x_3=1$. 
In the integrand, the variable $\vec{R}_{2\perp}$ turns into the integration variable $\vec{R^\prime }_{2\perp}$. 
Whereas, the kernels (\ref{W12}-\ref{W23}), after calculating the traces, depend on scalar products of the on-mass-shall four-momenta 
$k_{i}$, $k^\prime _{i}$, $p$, and $\omega$.   
Below we express these scalar products in terms of $x_i$ and the scalar products 
of $\vec{R}_{i\perp}$ and $\vec{R}_{j\perp}$.

By definition, we have
\begin{align*}
  &\omega\cd\omega=0,\,p\cd p=M^2,\,
  k_i\cd k_i=k^\prime _i\cd k^\prime _i=m^2,
  \\
  &\omega\cd k_i=x_i(\omega\cd p),\;\omega\cd k^\prime _i=x^\prime _i(\omega\cd p),\quad i=1,2,3.
\end{align*}
Note that the particle 3 is the spectator, so $k^\prime _3=k_3$. 

From $R_i^2=(k_i-x_ip)^2=-\vec{R}_{i\perp}^2$ we have:
\begin{align*}
 p\cd k_i=\frac{m^2+\vec{R}_{i\perp}^2}{2x_i}+\frac{1}{2}x_iM^2,
\end{align*}
and similarly for $p\cd k^\prime _i$. 

From $R_i\cd R_j=(k_i-x_ip)\cd(k_j-x_jp)=-\vec{R}_{i\perp}\cd\vec{R}_{j\perp}$ we obtain:
$$
k_i\cd k_j=-\vec{R}_{i\perp}\cd\vec{R}_{j\perp}+x_i(p\cd k_j)+x_j(p\cd k_i)-x_ix_jM^2,
$$
where $(p\cd k_i)$ are already found and similarly for the ``prime'' variables. In this way, we express all the four-vector scalar products in terms of the variables of the equations (\ref{gij}).


\end{document}